\documentclass[sigconf]{acmart}
\usepackage{multirow}
\usepackage{multirow}
\usepackage{amsmath}
\usepackage{xspace}
\usepackage{graphicx}     
\usepackage{subcaption}
\usepackage{enumitem}

\usepackage{algorithm}
\usepackage{algorithmicx}
\usepackage{algpseudocode}
\usepackage{threeparttable}
\usepackage{url}
\usepackage{xcolor}
\usepackage{float}

\makeatletter
\def\algbackskip{\hskip-\ALG@thistlm}
\makeatother

\AtBeginDocument{%
  }

\setcopyright{acmlicensed}
\copyrightyear{2027}
\acmYear{2027}
\acmDOI{XXXXXXX.XXXXXXX}
\acmConference[Conference acronym 'XX]{Make sure to enter the correct
  conference title from your rights confirmation email}{June 03--05,
  2027}{Woodstock, NY}
\acmISBN{978-1-4503-XXXX-X/2027/06}

\newcommand{\pname}{{ML-ECS}\xspace}

\begin{document}

\title{ML-ECS: Heterogeneity-Aware Collaborative Multimodal Learning for Edge–Cloud Synergies}

\author{Yuze Liu}
\affiliation{%
  \institution{Swinburne University of Technology}
  \city{Melbourne}
  \country{Australia}}
\email{yuzeliu@swin.edu.au}

\author{Shibo Chu}
\affiliation{%
  \institution{Tongji University}
  \city{Shanghai}
  \country{China}}
\email{2534148@tongji.edu.cn}

\author{Tiehua Zhang}
\authornote{Corresponding author.}
\affiliation{%
  \institution{Tongji University}
  \city{Shanghai}
  \country{China}}
\email{tiehuaz@tongji.edu.cn}

\author{Hao Zhou}
\affiliation{%
  \institution{Tongji University}
  \city{Shanghai}
  \country{China}}
\email{2353568@tongji.edu.cn}

\author{Zhishu Shen}
\affiliation{%
  \institution{Wuhan University of Technology}
  \city{Wuhan}
  \country{China}}
\email{z_shen@ieee.org}

\author{Jinze Wang}
\affiliation{%
  \institution{Swinburne University of Technology}
  \city{Melbourne}
  \country{Australia}}
\email{jinzewang@swin.edu.au}

\author{Jianzhong Qi}
\affiliation{%
  \institution{The University of Melbourne}
  \city{Melbourne}
  \country{Australia}}
\email{jianzhong.qi@unimelb.edu.au}

\author{Feng Xia}
\affiliation{%
  \institution{RMIT University}
  \city{Melbourne}
  \country{Australia}}
\email{feng.xia@rmit.edu.au}

\renewcommand{\shortauthors}{Trovato et al.}

\begin{abstract}
Edge–cloud synergies provide a promising paradigm for privacy-preserving deployment of foundation models in heterogeneous web environments, enabling lightweight edge models to support domain-specific multimodal we b services while cloud-hosted models facilitate collaborative knowledge sharing. However, in practical edge-cloud web environments, collaborative multimodal learning is challenged by modality heterogeneity (different modality combinations across domains) and model-structure heterogeneity (different modality-specific encoders/fusion modules). To address these issues, we propose ML-ECS, a collaborative multimodal learning framework that enables joint training between a server-based model and heterogeneous edge models. This framework consists of four components: (1) cross-modal contrastive learning (CCL) to align modality representations in a shared latent space, (2) adaptive multimodal tuning (AMT) to preserve domain-specific knowledge from local datasets, (3) modality-aware aggregation (MAA) to robustly aggregate while mitigating noise caused by missing modalities, and (4) SLM-enhanced CCL (SE-CCL) to facilitate bidirectional knowledge transfer between cloud and edge. Experimental results on various multimodal tasks show that \pname consistently outperform state-of-the-art baselines under varying modality availability, achieving improvements of 1.55\% to 16.75\% in Rouge-LSum and improving both client- and server-side performance. In addition, by communicating only low-rank LoRA parameters and fused representations, ML-ECS achieves high communication efficiency, requiring only 0.71\% of the total parameter volume.

\end{abstract}


\received{20 February 2007}
\received[revised]{12 March 2009}
\received[accepted]{5 June 2009}

\maketitle

\section{Introduction}
The rapid advancement of large language models (LLMs) has spurred growing interest in their deployment on mobile devices and in Web of Things (WoT) environments, where heterogeneous multimodal data from interconnected web-based platforms enable intelligent applications such as mobile agentic assistants~\cite{li2026curiosity}, smart wearable devices~\cite{yang2026egocentric}, and health monitoring systems~\cite{zhu2025heal}. LLMs are typically deployed on remote server clusters as foundation models (FMs) to accommodate the substantial computational demands of inference~\cite{hao2024hybrid,liu2025grl,jiang2026divide} and facilitate fine-tuning on multimodal data collected from diverse sources, thereby enhancing their adaptability to diverse multimodal tasks~\cite{zhan2024anygpt}. However, the massive scale of these models and their dependence on large, heterogeneous training data corpora pose practical limitations, including data-privacy risks, substantial computational/energy costs, and limited domain adaptability without further fine-tuning~\cite{thirunavukarasu2023large,hao2024hybrid}. These limitations motivate edge-cloud synergistic learning frameworks, where lightweight edge models inherit transferable capabilities from cloud-hosted FMs to serve domain-specific mining workloads close to data sources, while resource-intensive FMs remain in the cloud to orchestrate collaboration and continuously improve the performance of distributed edge models~\cite{yang2025fed,nguyen2025device}.

Recent studies have investigated edge–cloud collaborative inference~\cite{yang2025ec2moe,wang2024cloud}, in which lightweight data preprocessing is performed on edge devices while computation-intensive inference is offloaded to the cloud, thereby balancing efficiency and accuracy across heterogeneous resources in edge–cloud environments. However, these approaches fail to consider continual training to further enhance model capability. By contrast, another line of work focuses on personalized federated learning (PFL) ~\cite{li2020federated,zhang2023adaptive} as an edge–cloud solution to mitigate statistical heterogeneity, particularly non-IID data distributions that manifest as domain-specific biases in edge environments during the collaborative learning process. These methods are constrained by their limited consideration of potential collaboration between cloud-based FMs and domain-specific edge models~\cite{chen2025survey,deng2025crosslm,liu2025structure}.

Furthermore, developing effective edge–cloud synergistic learning in heterogeneous web environments still faces two key challenges:
\begin{itemize}[leftmargin=0.3cm, noitemsep]
    \item \textbf{\textit{Modality Heterogeneity}}: In practical edge–cloud web environments, edge models are deployed to support domain-specific web services, and variations across domains naturally lead to heterogeneous combinations of data modalities. For example, in emergency response scenarios (e.g., earthquakes or floods), effective analysis requires integrating multimodal data such as images or videos (on-site photos), text (social media reports), and audio signals (distress calls)~\cite{alam2018crisismmd,wang2021footprint}. In contrast, web-connected healthcare services process different combinations of modalities, such as electronic health records, physiological signals (e.g., ECG and EEG recordings), and medical images (e.g., CT scans)~\cite{lau2018dataset}. Such domain-induced modality heterogeneity across web-enabled edge services can significantly degrade collaborative learning performance~\cite{feng2023fedmultimodal}.
    \begin{sloppypar}
          \item \textbf{\textit{Model Structure Heterogeneity}}: To accommodate domain-specific modality heterogeneity, edge models are often configured with distinct model architectures. These architectural differences arise from the use of different combinations of modality-specific feature extractors or fusion modules according to the modalities available on each edge device, resulting in model structure heterogeneity across edge models~\cite{wang2025mitigating}. Conventional homogeneous federated learning methods, which assume identical model architectures, are unable to conduct collaborative training under such heterogeneous model structures~\cite{tan2022towards}.  
    \end{sloppypar}

\end{itemize}

To address these challenges, we propose \pname, a collaborative multimodal learning framework for edge-cloud synergy that enables joint training between a server-side model and domain-specific models deployed across edge devices. The framework comprises four key components: (i) cross-modal contrastive learning (CCL) to effectively encode modality representations for edge models; (ii) adaptive multimodal tuning (AMT) of on-device models to capture domain-specific biases within local private datasets; (iii) modality-aware aggregation (MAA); and (iv) SLM-enhanced cross-modal contrastive learning (SE-CCL) for updating the server-based model.

In summary, the main contributions of this work are:
\begin{itemize}[leftmargin=0.3cm, noitemsep]
    \item We propose \pname, a novel collaborative framework that enables synergistic model learning across edge-cloud consortia through four key modules: CCL, AMT, MAA, and SE-CCL. In \pname, modality representations are effectively encoded into a unified latent space across edge devices through the CCL process, in which a contrastive alignment strategy is designed to capture multimodal semantic content. Meanwhile, AMT of edge models helps preserve domain-specific multimodal knowledge during the collaborative learning process.
    \item We design an MAA mechanism to mitigate noise in the parameters uploaded by on-device edge models caused by modality heterogeneity, and introduce SE-CCL to enable effective bidirectional knowledge transfer between the server-based model and on-device models.
    \item We conduct extensive experiments on two downstream tasks, namely summary generation and classification, using the VAST and UR-FALL datasets, respectively, and demonstrate that \pname consistently outperforms baseline methods. In addition, we provide a detailed analysis of the communication overhead. Experimental results demonstrate that the proposed \pname significantly outperforms state-of-the-art baselines on both tasks while maintaining high communication efficiency throughout the collaborative learning process.
    \item We have made our source code publicly available at https://\\github.com/papercode-DFL/ML-ECS to encourage further research and reproducibility.
\end{itemize}
\section{Related Work}

Current LLM deployment remains largely cloud-centric. Cloud servers offer the computing and memory resources needed for large-scale models. This paradigm often leads to network-dependent latency, higher energy consumption, and privacy concerns. In contrast, running a full-size LLM on edge devices is generally impractical, given limited computing and memory resources available on edge devices. To address these constraints, recent studies have explored edge–cloud collaborative serving and learning for LLM-powered applications. For example, 
CSCA~\cite{sun2025edge} proposes an edge cognitive semantic communication agent that uses a large model for modality alignment and intent understanding, and generates personalized wireless communication policies to improve semantic accuracy. Co-PLMs~\cite{liu2025structure} further advances collaborative learning by introducing a co-tuning framework for large and small language models, using structure-agnostic mutual learning to exchange knowledge across edge models with heterogeneous architectures. 

Several studies use federated learning (FL) to mitigate performance degradation caused by missing modalities. For example, 
FedMLLM~\cite{xu2024fedmllm} addresses bias introduced by multimodal heterogeneity via two simple yet effective modality-agnostic strategies, including improved prompting and regularization. Fedilora~\cite{yang2025fedilora} studies federated multimodal fine-tuning under heterogeneous LoRA ranks and missing modalities by combining dimension-wise aggregation with lightweight layer-wise model editing. 
However, existing methods largely overlook the potential collaboration between on-device SLMs and server-side LLMs in multimodal edge-cloud synergy settings. As a result, these methods are limited in their ability to handle modality heterogeneity in edge–cloud environments, whereas this challenge is addressed by our proposed framework.


\section{Preliminaries}
\begin{figure}[t!]
    \vspace{-2mm}
    \centering
    \includegraphics[width=0.85\linewidth]{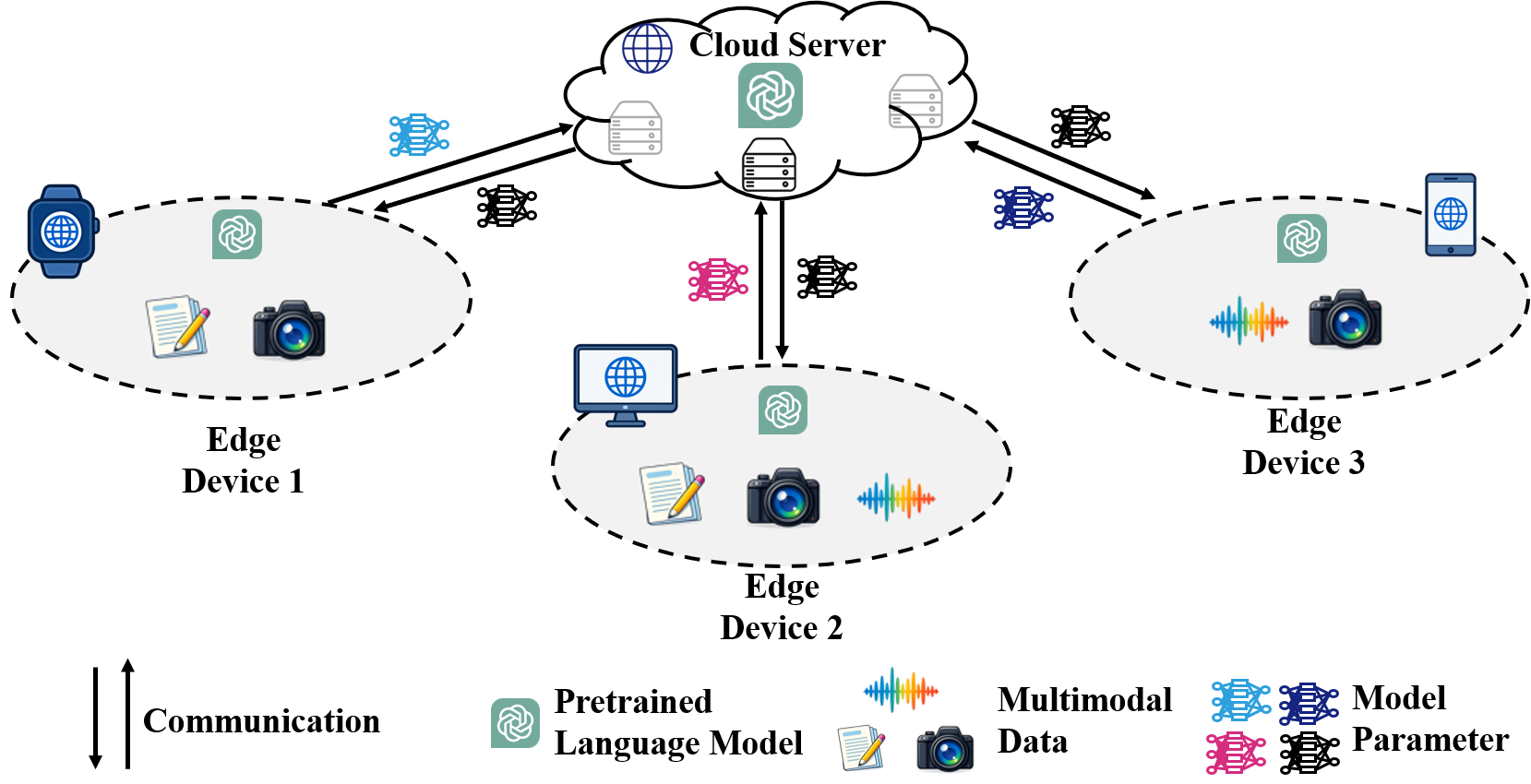}
        \vspace{-4mm}
  \caption{General architecture of edge-cloud synergies.}
  \label{fig:1}
      \vspace{-8mm}
\end{figure}
\noindent\textbf{LLMs for Multimodal Input}\quad
LLMs are typically adapted to multimodal data through a unified modeling paradigm consisting of a multimodal feature extractor, a connector module, and a text-based language model backbone~\cite{yin2024survey}. Formally, given a set of multimodal input $\{\emph{x}(m_i)\}$, where a modality $m_i\in\mathcal{M}$, a multimodal feature extractor $\emph{E}$ encodes each input into a modality-specific feature vector $\emph{z}(m_i) = \emph{E}^{m}_i(\emph{x}(m_i))$ using the corresponding encoder $\emph{E}^{m}_i$. Specifically, BERT~\cite{devlin2019bert} is used for the text modality, CLIP~\cite{radford2021learning} for the image modality, LSTM~\cite{hochreiter1997long} for the signal modality, and BEATS~\cite{chen2022beats} for the audio modality. The feature vectors $\emph{z} = \{\emph{z}(m_i)\}, m_i\in\mathcal{M}$ are subsequently mapped into a shared latent space via a connector module $\emph{C}$ and then utilized by a downstream text-based language model backbone $\emph{B}$, enabling effective integration of multimodal information. 


\noindent\textbf{Problem Definition}\quad
As illustrated in \figurename~\ref{fig:1}, we consider an edge-cloud web environment comprising a central server and a set of edge devices, denoted by $\mathcal{D} = \{d_j\}_{j=1}^{N}$, where $N$ represents the total number of devices. Each device owns a private dataset with heterogeneous modalities, denoted as $D = {\emph{x}(m_i)}_{i=1}^{|\mathcal{M}(D)|}$, where $\mathcal{M}(D) \subseteq \mathcal{M}$ represents the set of modalities available on the device. The orderd list of available modality names on edge device $i$ is denoted by $[\mathcal{M}_i]$, where $i \in [1,N]$. For example, if edge device $i$ has access to text and image data, then $[\mathcal{M}_i] = [\text{Text}, \text{Image}]$. Let $D' = {\emph{x}(m_i)}_{i=1}^{|\mathcal{M}|}$ denote a omni-modal public dataset accessible to the server and all devices. The corresponding collection of public datasets with heterogeneous modalities for the devices is denoted by $\{{D'}\}_{j=1}^{N}$. To effectively leverage multimodal information, the local language models are extended to a unified model $\emph{M} = \{\emph{E}, \emph{C}, \emph{B}\}$, which consists of a multimodal feature extractor $\emph{E}$, a connector module $\emph{C}$, and a text-based language model backbone $\emph{B}$. This unified architecture is deployed across both devices and the central server, where the on-device language models correspond to SLMs defined as $\emph{B}^{d}$, and the server-side language model corresponds to an LLM defined as $\emph{B}^{s}$.

The server and edge devices collaboratively seek to enhance the performance of the unified on-device models $\{\emph{M}^{d}\}_{j=1}^{N}$ and the server-side model $\emph{M}^{s}$ through collaborative learning, while preserving the privacy of local data. The overall objective function to be minimized over the collaborative learning process is defined as:

\begin{equation}
    \min_{\theta(\emph{M}^s),\{\theta(\emph{M}_{j}^{d})\}_{j = 1}^{N}}\!\!\!\mathcal{L}(\theta(\emph{M}^{s}), \{\theta(\emph{M}_{j}^{d})\}_{j=1}^{N};D^{'}, \{D_{j}\}_{j=1}^{N},\{D^{'}_{j}\}_{j=1}^{N})
\end{equation}
where $\theta(\emph{M}^s)$ and $\{\theta(\emph{M}^{d}_{j})\}_{j=1}^{N}$ denote the trainable parameters of the server and the devices, respectively. Furthermore, we consider multimodal heterogeneity across devices, denoted by $\exists j\in[1,...,N], |\mathcal{M}(D_{j})| < |\mathcal{M}|$, where $|\mathcal{M}(D_{j})|$ denotes the number of modalities available on device $j$.

\section{Methodology}

\begin{figure*}[t!]
    \centering
    \vspace{-2mm}
\scalebox{1.1765}[1]{%
        \includegraphics[width=0.78\linewidth]{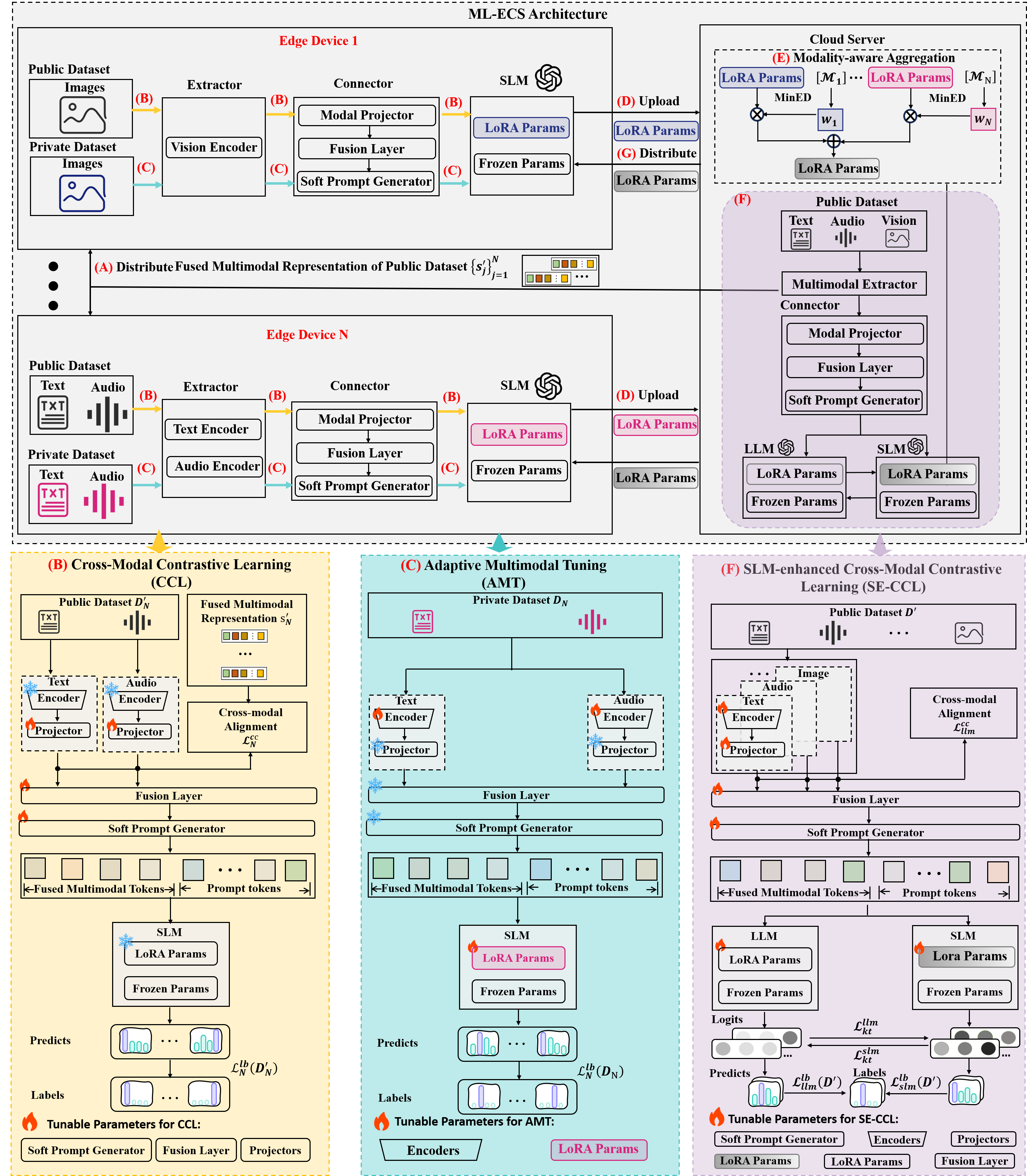}
    }
    \vspace{-3mm}
  \caption{Overview of our proposed ML-ECS.}
  \label{fig:2}
  \vspace{-5mm}
\end{figure*}
\figurename~\ref{fig:2} illustrates an overview of the \pname architecture. We first introduce cross-modal contrastive learning (CCL), which enhances the encoding of modality representations across devices by minimizing a cross-modal alignment loss computed from server-provided fused multimodal vectors. Next, adaptive multimodal tuning (AMT) is performed on devices to capture heterogeneous multimodal bias. On the server side, the server-based SLM updates its model using a modality-aware aggregation (MAA) mechanism to mitigate performance degradation caused by modality heterogeneity. Furthermore, SLM-enhanced cross-modal contrastive learning (SE-CCL) is conducted between the server-based unified model and the server-side SLM to facilitate bidirectional knowledge exchange. Finally, the server distributes the tunable parameters of its SLM to the devices to propagate knowledge throughout the edge-cloud environment.
\subsection{Cross-modal Contrastive Learning}
As illustrated in the upper part of \figurename~\ref{fig:2}, the unified models deployed across edge devices, denoted by $\{\emph{M}^d_j\}_{j=1}^{N}$, together with the server-side unified model $\emph{M}^s$ share a common architecture comprising a multimodal feature extractor $\emph{E}$, a connector module $\emph{C}$, and a text-based language model backbone $\emph{B}$. Specifically, the connector module $\emph{C}$ consists of a modality projector, a fusion layer, and a soft prompt generator. For edge device $j$, the feature of modality $m_i$, where $i \in[1,... ,|\mathcal{M}(D_j)|]$, is denoted by $\emph{z}_j(m_i)$ and is generated by the corresponding modality-specific encoder $\emph{E}^{m}_i$. For example, an audio encoder extracts audio features from audio data, as shown in Part (B) of \figurename~\ref{fig:2}. Then, the set of projectors within the connector module is designed to map modality-specific features into a shared latent space, defined as follows:
\begin{equation}
    \emph{h}_j(m_i) = f_{j}^{p}(\emph{z}_j(m_i))
\end{equation}
where $\emph{h}_j(m_i), i \in[1,... ,|\mathcal{M}(D_j)|]$ denotes the representation of modality $m_i$ on device $j$. 

However, in real-world edge-cloud multimodal scenarios, modality heterogeneity, reflected in variations in both the number and combinations of modalities available across edge devices, hinders effective multimodal representation learning and consequently degrades downstream task performance during collaborative learning.~\cite{che2024leveraging,wang2024fedmmr}. Previous works primarily focus on pairwise modality alignment based on cosine similarity~\cite{zhu2023languagebind} or adopt suboptimal neural fusion strategies for integrating more than two modalities~\cite{liu2022umt,xu2023mplug}. Although these multimodal alignment strategies yield  performance gains under standalone training settings, they struggle with effectively encoding multimodal representations into a unified latent space shared across all edge devices, particularly in the presence of modality heterogeneity. This limits their practical utility in multimodal edge–cloud environments.

To address these challenges, we design a contrastive cross-modal alignment strategy that encourages modality representations, which share the same semantic content, to be close to one another, while pushing semantically uncorrelated representations farther apart. As a result, we learn a better aligned and semantically meaningful multimodal embedding space. 
To quantify the semantic alignment among multimodal representations, 
we employ the notion of vector volume in a hyperdimensional space, computed as the square root of the determinant of the corresponding Gram matrix~\cite{gantmacher1959matrix}.  Mathematically, let $\{\emph{v}_i\}_{i=1}^{k}$, where $\emph{v}_{i}\in\mathcal{R}^n$, denote a set of vectors, which can be arranged as the columns of a matrix $\emph{A}\in\mathcal{R}^{n\times k}$. The corresponding Gram matrix $\mathbf{G}$ is defined as:
\begin{equation}
    \mathbf{G}({\{\emph{v}_i\}_{i=1}^{k}}) = \emph{A}^{T}\cdot\emph{A}
\end{equation}

The vector volume can be computed from the Gram matrix as follows:
\begin{equation}
    \mathbf{V}({\{\emph{v}_i\}_{i=1}^{k}}) = \sqrt{\emph{det}(\mathbf{G}({\{\emph{v}_i\}_{i=1}^{k}}))}
    \label{eq:1}
\end{equation}
where $\emph{det}(\cdot)$ denotes the determinant of the matrix. Note that $\emph{det}(\mathbf{G}) = \emph{det}(\emph{A}^{T}\cdot\emph{A}) = (\emph{det}(\emph{A}))^2 \geq 0$, which ensures that the square root in Eq.~\ref{eq:1} is well defined. Vector volume provides an intuitive geometric perspective for measuring the degree of alignment among vectors~\cite{cicchettigramian} and can be naturally applied to multimodal representations. The volume spanned by a set of modality representations quantifies their geometric dispersion in a hyperdimensional space. A smaller volume indicates that the modality representations are closely aligned, suggesting strong semantic consistency among the multimodal inputs. Conversely, a larger volume implies greater dispersion, indicating semantic divergence among the modality representations. Furthermore, the vector volume is inherently scalable and can be applied to any number of modalities, ranging from $2$ to $n$, thus providing a viable solution to modality heterogeneity in edge–cloud environments.

\noindent\textbf{Cross-modal Contrastive Loss} We propose a novel contrastive loss $\mathcal{L}^{cc}$ that leverages the vector volume defined in Eq.~\ref{eq:1}. Specifically, one modality representation is selected as the anchor for a given sample, with the remaining modality representations from the same sample serving as positive samples, while representations from different samples are used as negative samples via a negative sampling strategy~\cite{radford2021learning}. The loss computation process is defined as follows:
\begin{equation}
    \mathcal{L}^{O2A}_j\!=\!-\log\frac{\exp(-\mathbf{V}(\{\emph{h}_{j,v}^{A}(m^A)\}\cup\{\emph{h}_{j,v}(m)\}_{m\in\mathcal{M}_{j}^{O}}))}{\sum_{u =1}^{U}\exp(-\mathbf{V}(\{\emph{h}_{j,v}^{A}(m^A)\}\cup\{\emph{h}_{j,u}(m)\}_{m\in\mathcal{M}_{j}^{O}}))}
\end{equation}

\begin{equation}
    \mathcal{L}^{A2O}_j\!=\!-\log\frac{\exp(-\mathbf{V}(\{\emph{h}_{j,v}^{A}(m^A)\}\cup\{\emph{h}_{j,v}(m)\}_{m\in\mathcal{M}_{j}^{O}}))}{\sum_{u =1}^{U}\exp(-\mathbf{V}(\{\emph{h}_{j,u}^{A}(m^A)\}\cup\{\emph{h}_{j,v}(m)\}_{m\in\mathcal{M}_{j}^{O}}))}
\end{equation}
where $U$ denotes the number of the negative sample, and $\mathcal{M}_j^{O}\subseteq\mathcal{M}$ denotes the set of non-anchor modality representations for edge device $j$. The overall cross-modal contrastive loss $\mathcal{L}^{cc}_j$ for device $j$ is formulated as follows: $\mathcal{L}^{cc}_j = \frac{1}{2} (\mathcal{L}^{A2O}_{j}+\mathcal{L}^{O2A}_{j})$. Note that the server-provided fused omni-modal representations $\emph{s}^{'}$ on the public dataset are selected as anchors during the CCL process on devices, since the public dataset is accessible to both the server and all edge devices, and the server is equipped with an omni-modal extractor that enables more effective representation learning from omni-modal inputs on the public dataset.

\noindent\textbf{Soft Prompt Tuning} To enable the language model backbone $\emph{B}$ to fully leverage multimodal content, a fusion layer, implemented using a multi-layer perceptron (MLP), is designed to generate a fused multimodal representation. The fusion process can be defined as follows:
\begin{equation}
    \emph{s}_{j} = f_{u}(\{\emph{h}_{j}(m_{i})\}_{i = 1}^{|\mathcal{M}_j|})
    \label{eq:fused}
\end{equation}
where $j\in[1, ..., N]$. Subsequently, the fused multimodal representation is processed by a soft prompt generator to adapt the input to the language model backbone $\emph{B}_{j}$, together with the tokenized prompt template containing the user instruction, produced by the tokenizer of the local SLM on each device $\mathbf{tokenizer}^{slm}$, as defined as follows:
\begin{equation}
    \emph{p}_{j} = f_{con}(f_{spg}(\emph{s}_{j}), \mathbf{tokenizer}^{slm}_j(\textit{prompt}))) 
\end{equation}

In this work, we employ a two-layer MLP with a GeLU activation function as the soft prompt generator~\cite{yang2023prompt}. It generates fused multimodal tokens from the fused multimodal representation. This token generation process is denoted by $f_{spg}$. $f_{con}$ denotes the concatenation operation, which places the fused multimodal tokens before the prompt tokens, as illustrated in \figurename~\ref{fig:2} Part (B). Furthermore, the overall loss function for the CCL process can be formulated as follows:
\begin{equation}
    \mathcal{L}_j^{ccl} (D^{'}_j) = \mathcal{L}_j^{lb} (D^{'}_j) + \mathcal{L}_j^{cc}
    \label{eq:loss_ccl}
\end{equation}
where $\mathcal{L}_j^{lb}(D^{'}_j)$ denotes the supervised finetuning loss on the public dataset for device $j$, enabling it to enhance multimodal representation encoding under the guidance of server-provided omni-modal content from the public dataset. The process of CCL is defined as $f_{ccl}(\emph{M}^{d}_j;D^{'}_j)$.

\subsection{Adaptive Multimodal Tuning}
To mitigate performance degradation caused by the divergence between on-device multimodality and server-side omni-modality, we propose an AMT process that adaptively captures modality-dependent knowledge from the local private datasets of devices during the collaborative learning process. As illustrated in Part (C) of \figurename~\ref{fig:2}, we employ low-rank adaptation (LoRA)~\cite{hu2022lora} to fine-tune the SLM backbone, with its trainable parameters denoted by $\phi_{lora}(\emph{B}^d)$, while jointly optimizing the feature extraction module. The loss function for this process can be formulated as follows: $\mathcal{L}_{j}^{amt}(D_j)= \mathcal{L}_j^{lb}(D_j)$, where $\mathcal{L}_j^{lb}(D_j)$ denotes the supervised finetuning loss on the local private dataset $D_j, j\in[1,...,N]$. The process of AMT is defined as $f_{amt}(\emph{M}_{j}^{d};D_j)$.

\subsection{Modality-aware Aggregation}
Upon completion of on-device training, each device uploads its LoRA parameter $\phi_{lora}(\emph{B}^d)$ of the on-device SLM backbone, along with its ordered list of available modality names $[\mathcal{M}_j], j\in[1,...,N]$ to the server. To alleviate noise in the uploaded parameters arising from divergence across on-device modality heterogeneity~\cite{chen2024fedmbridge}, we propose an MAA mechanism that evaluates the importance of LoRA parameters uploaded by devices. Due to variations in the number of available modalities across edge devices, on-device models trained on modality combinations that differ substantially from those of neighboring devices tend to exhibit greater variance and are therefore assigned lower aggregation weights~\cite{nguyen2024fedmac}. We leverage minimum edit distance (MinED)~\cite{marzal1993computation} as an aggregation-weight metric to identify informative model parameters from edge devices. A larger sum of pairwise MinED scores among neighboring edge devices indicates greater modality diversity. The normalized aggregation weight assigned to the LoRA parameters uploaded by edge device $j$, based on the lists of available modality names across edge devices, is defined as follows: 
\begin{equation}
    w_j = \frac{\sum_{i\in[1,N], i\neq j}(1/(\text{MinED}([\mathcal{M}_j], [\mathcal{M}_i])+1))}{\sum_{j\in[1,N]}\sum_{i\in[1,N], i\neq j}(1/(\text{MinED}([\mathcal{M}_j], [\mathcal{M}_i])+1)}
    \label{eq:weight}
\end{equation}

After the aggregation weights are assigned, the aggregated LoRA parameters are applied to the server-side SLM backbone $\emph{B}^s_{slm}$. The aggregation process can be defined as $f_{maa}(\{\phi(\emph{B}_{j}^{d})\}_{j = 1}^{N})$.

\subsection{SLM-enhanced Cross-modal Contrastive Learning}
To enable collaborative learning between heterogeneous model architectures on the server-side SLM backbone and the server-based unified model, we propose SE-CCL to facilitate bidirectional knowledge transfer between devices and the server. As illustrated in Part (F) of the \figurename~\ref{fig:2}, compared to the on-device CCL, the server-side SE-CCL exhibits two key differences. First, owing to the omni-modality of the public dataset on the server, one modality from the omni-modal set is randomly selected as the anchor modality for computing the cross-modal contrastive loss. Second, SE-CCL comprises pairs of a server-based SLM backbone and a LLM backbone, which exchange knowledge through a knowledge-transfer loss applied to their normalized output logits. We employ a pooling-based knowledge transfer loss~\cite{liu2025structure} to address divergence singularities~\cite{cover1999elements} caused by sparse output distributions in language models. This loss can be formulated as follows:
\begin{equation}
    \mathcal{L}_{kt}(\emph{Y}^{slm},\emph{Y}^{llm}) = \sum_{i = 1}^{S}KLD(\emph{y}_i^{slm},\emph{y}_i^{llm}) 
\end{equation}
$S = \min(S_1,S_2)$, where $S_1$ and $S_2$ denote the sequence lengths of the pooled output logits $\emph{Y}^{slm}$ and $\emph{Y}^{llm}$, respectively. During the process of SE-CCL, the overall loss function for the server-based unified model can be formulated as follows:
\begin{equation}
    \mathcal{L}_{llm}^{se}(D^{'})= \mathcal{L}^{ccl}_{llm}(D^{'}) + \sum_{j = 1}^{|D^{'}|} \mathcal{L}_{kt}^{llm}(\emph{Y}_{j}^{slm},\emph{Y}_{j}^{llm})
\end{equation}
where the loss function for the CCL process, denoted as $\mathcal{L}_{llm}^{ccl}(D^{'})$, is defined in Eq.\ref{eq:loss_ccl}. The loss function for server-based SLM is defined as follows:
\begin{equation}
    \mathcal{L}^{se}_{slm}(D^{'}) = \mathcal{L}^{lb}_{slm}(D^{'}) + \sum_{j = 1}^{|D^{'}|} \mathcal{L}_{kt}^{slm}(\emph{Y}_{j}^{llm},\emph{Y}_{j}^{slm})
\end{equation}

The SE-CCL process can be formulated as $f_{se}(\emph{M}^s, \emph{B}^s_{slm})$.

\subsection{Overall Training Process}
The workflow of \pname proceeds as follows, and the corresponding pseudocode is provided in \textbf{Appendix A}.
\begin{enumerate}[leftmargin=0.3cm, noitemsep]
    \item In the $t$-th communication round, fused omnimodal representations are generated by the server-based unified model and distributed to each device.
    \item Each edge device sequentially performs CCL and AMT, then uploads the trainable LoRA parameters of its SLM backbone, along with its ordered list of available modalities, to the central server.
    \item The server aggregates the uploaded LoRA parameters using the MAA mechanism to update its server-based SLM and subsequently performs SE-CCL. The updated LoRA parameters of the server-based SLM are then distributed to each device.
    \item Each device receives the LoRA parameters from the server and updates its local SLM backbone accordingly.
\end{enumerate}
\section{Experiment}
In this section, we conduct extensive experiments to evaluate the effectiveness of our proposed framework from different perspectives. Specifically, we aim to answer the following research questions: \textbf{RQ1:} How does \pname compare with the five baselines in terms of device-side and server-side performance? \textbf{RQ2:} How does \pname compare with the five baselines in terms of communication cost? \textbf{RQ3:} How does the number of edge devices affect the scalability of \pname in edge-cloud synergies? \textbf{RQ4:} How does the size of the edge model affect the performance and communication cost of \pname? \textbf{RQ5:} How do the key components, including CCL, MAA, and SE-CCL, contribute to the performance of \pname?
\subsection{Experiment Setup}

We assume an edge-cloud web system comprising one cloud server and $N = 3$ edge devices. We deploy MiniLLM-GPT2-720M~\cite{gu2023minillm} as the on-device SLM backbones, while the cloud server uses GPT-J-6B~\cite{wang2021gpt} for server-based LLM backbone.

\noindent\textbf{Datasets}\quad 
We evaluate our framework on two multi-modal datasets with different tasks: (1) VAST~\cite{chen2023vast} focusing on summary generation, is a large-scale omni-modality dataset containing 27M video clips with aligned visual, audio, and subtitle modalities. We randomly select a subset of 9,680 clips for experiments. (2) UR\text{-}FALL~\cite{kwolek2014human}, designed for classification, comprises 11,544 samples extracted from 70 RGB-depth video sequences. 

\noindent\textbf{Data Partition}\quad 
To simulate modality heterogeneity, we assume the availability of each modality $m\in\mathcal{M}$ follows an independent Bernoulli distribution $\text{Bernoulli} (\rho_m)$. Here, we use a \textbf{modality existing rate (MER)} $\rho_m$ to indicate the probability that an edge device possesses modality $m$. Specifically, a lower MER indicates greater heterogeneity in the number of modalities available across edge devices.

\noindent\textbf{Baselines}\quad 
We compare \pname with the following baselines: Standalone, Multi-FedAvg, FedMLLM~\cite{xu2024fedmllm}, Fedilora~\cite{yang2025fedilora}, and Co-PLMs~\cite{liu2025structure}. The corresponding implementation details are provided in \textbf{Appendix B}.

\noindent\textbf{Evaluation Metrics}\quad
For the summarization task, we use \textbf{Rouge-LSum (RLS)} to measure lexical overlap and \textbf{BERTScore (BS)} to assess semantic similarity. For each metric, we report the average (\textbf{Avg. RLS} and \textbf{Avg. BS}), best (\textbf{B. RLS} and \textbf{B. BS}), and worst (\textbf{W. RLS} and \textbf{W. BS}) scores across all edge-device models. For the classification task, we use the \textbf{F1-score (F1)}, and similarly report \textbf{Avg. F1}, \textbf{B. F1}, and \textbf{W. F1} across edge-device models.

\subsection{Performance Analysis (RQ1)}
Following the settings adopted in previous studies, we simulate varying levels of modality heterogeneity in edge-cloud synergies by setting the MER to $\rho_m \in \{0.5, 0.7, 0.8\}$~\cite{jin2023rethinking}. A smaller value of $\rho_m$ indicates greater variation in modality distributions across edge devices, resulting in more severe modality heterogeneity. 
Table~\ref{tab:summary} summarizes the performance of \pname and five baselines, including Standalone, Multi-Fed-Avg, FedMLLM, Fedilora, and Co-PLMs. Since Multi-Fed-Avg, FedMLLM, and Fedilora do not maintain a server-side model, their server-side results are marked as ``--''. All experiments are conducted five times with different random seeds, and the average results are reported.

As shown in Table~\ref{tab:summary}, \pname consistently outperforms the baselines across different MER settings in terms of Avg. BS, Avg. RLS, BS and RLS on VAST, as well as Avg. F1 and F1 on UR-FALL. Specifically, on the VAST dataset, it achieves improvements of 1.98\%--10.98\%, 1.55\%--16.75\%, 1.31\%--4.03\% and 1.67\%--5.19\% in Avg. BS, Avg. RLS, BS and RLS, respectively. 
On average, \pname achieves gains of 
6.38\%, 2.94\%, 7.85\%, and 1.33\% in Avg. F1, B. F1, W. F1, and F1, respectively, on the UR-FALL dataset. Compared with Standalone, Multi-Fed-Avg, FedMLLM, Fedilora, and Co-PLMs, \pname improves the device-side Avg. BS on the VAST dataset by 6.02\%, 7.00\%, 4.32\%, 4.11\% and 3.88\%, respectively, and the device-side Avg. F1 on the UR-FALL dataset by 7.55\%, 11.33\%, 5.48\%, 3.44\%, and 4.11\%, respectively. These results demonstrate the effectiveness of \pname in learning more informative multimodal representations for device-side models in multimodal edge-cloud synergy settings. The proposed framework also consistently improves server-side performance. Compared with Standalone and Co-PLMs, \pname improves the server-side BS on VAST by 2.43\% and 2.37\%,  and the server-side RLS by 2.33\% and 4.21\%. 
It is worth noting that the performance of all methods decreases as $\rho$ declines from $0.8$ to $0.5$, with average reductions of 7.87\% in Avg. BS and 10.75\% in Avg. RLS on VAST, as well as 11.83\% in Avg. F1 on UR-FALL. This degradation is attributable to the increased modality heterogeneity associated with a lower MER, which renders collaborative multimodal learning more challenging. Nevertheless, \pname exhibits greater robustness than the baselines. Specifically, its Avg. BS and Avg. RLS on VAST decrease by only 5.18\% and 6.87\%, respectively, while its Avg. F1 on UR-FALL declines by only 5.87\%. 
These results demonstrate the effectiveness of \pname in mitigating the performance degradation of modality heterogeneity in edge-cloud synergies.
\begin{table*}[]
\vspace{-2mm}
\caption{Comparison of \pname on the VAST and UR-FALL datasets. The best results are highlighted in bold.}
\vspace{-2mm}
\label{tab:summary}
\begin{threeparttable}
\resizebox{0.9\linewidth}{!}{
\begin{tabular}{cc|cccccccc|cccc}
\hline
\multirow{3}{*}{MER}                       & \multirow{3}{*}{Method} & \multicolumn{8}{c|}{VAST}                                                                                                                                    & \multicolumn{4}{c}{UR-FALL}                                                            \\ \cline{3-14} 
                                           &                         & \multicolumn{6}{c|}{Client Perf.}                                                                               & \multicolumn{2}{c|}{Server Perf.} & \multicolumn{3}{c|}{Client Perf.}                                     & Server Perf.   \\ \cline{3-14} 
                                           &                         & Avg. BS        & B. BS          & W. BS          & Avg. RLS         & B. RLS         & \multicolumn{1}{c|}{W. RLS}         & BS              & RLS             & Avg. F1        & B. F1          & \multicolumn{1}{c|}{W. F1}          & F1             \\ \hline
\multirow{6}{*}{$\rho_m = 0.5$} & Standalone              & 0.272          & 0.279          & 0.260          & 0.218          & 0.226          & \multicolumn{1}{c|}{0.207}          & 0.382           & 0.299           & 0.697          & 0.768          & \multicolumn{1}{c|}{0.647}          & 0.912          \\
                                           & Multi-Fed-Avg           & 0.264          & 0.278          & 0.252          & 0.209          & 0.218          & \multicolumn{1}{c|}{0.202}          & -               & -               & 0.614          & 0.749          & \multicolumn{1}{c|}{0.527}          & -              \\
                                           & FedMLLM                 & 0.276          & 0.281          & 0.268          & 0.231          & 0.234          & \multicolumn{1}{c|}{0.222}          & -               & -               & 0.672          & 0.771          & \multicolumn{1}{c|}{0.594}          & -              \\
                                           & Fedilora                & 0.274          & 0.278          & 0.262          & 0.228          & 0.236          & \multicolumn{1}{c|}{0.212}          & -               & -               & 0.704          & 0.759          & \multicolumn{1}{c|}{0.567}          & -              \\
                                           & Co-PLMs                 & 0.274          & 0.283          & 0.259          & 0.228          & 0.239          & \multicolumn{1}{c|}{0.214}          & 0.372           & 0.289           & 0.686          & 0.764          & \multicolumn{1}{c|}{0.617}          & 0.901          \\
                                           & \textbf{Ours}           & \textbf{0.293} & \textbf{0.301} & \textbf{0.284} & \textbf{0.244} & \textbf{0.249} & \multicolumn{1}{c|}{\textbf{0.232}} & \textbf{0.387}  & \textbf{0.304}  & \textbf{0.754} & \textbf{0.791} & \multicolumn{1}{c|}{\textbf{0.698}} & \textbf{0.918} \\ \hline
\multirow{6}{*}{$\rho_m = 0.7$} & Standalone              & 0.287          & 0.292          & 0.278          & 0.234          & 0.239          & \multicolumn{1}{c|}{0.225}          & 0.383           & 0.300           & 0.729          & 0.774          & \multicolumn{1}{c|}{0.713}          & 0.917          \\
                                           & Multi-Fed-Avg           & 0.284          & 0.297          & 0.276          & 0.222          & 0.228          & \multicolumn{1}{c|}{0.216}          & -               & -               & 0.746          & 0.794          & \multicolumn{1}{c|}{0.684}          & -              \\
                                           & FedMLLM                 & 0.294          & 0.301          & 0.281          & 0.231          & 0.233          & \multicolumn{1}{c|}{0.225}          & -               & -               & 0.772          & 0.806          & \multicolumn{1}{c|}{0.708}          & -              \\
                                           & Fedilora                & 0.294          & 0.302          & 0.279          & 0.233          & 0.242          & \multicolumn{1}{c|}{0.223}          & -               & -               & 0.768          & 0.794          & \multicolumn{1}{c|}{0.724}          & -              \\
                                           & Co-PLMs                 & 0.292          & 0.298          & \textbf{0.281} & 0.241          & 0.246          & \multicolumn{1}{c|}{0.237}          & 0.387           & 0.292           & 0.774          & 0.801          & \multicolumn{1}{c|}{0.729}          & 0.912          \\
                                           & \textbf{Ours}           & \textbf{0.300} & \textbf{0.307} & 0.280          & \textbf{0.249} & \textbf{0.252} & \multicolumn{1}{c|}{\textbf{0.242}} & \textbf{0.393}  & \textbf{0.307}  & \textbf{0.786} & \textbf{0.813} & \multicolumn{1}{c|}{\textbf{0.734}} & \textbf{0.928} \\ \hline
\multirow{6}{*}{$\rho_m = 0.8$} & Standalone              & 0.292          & 0.296          & 0.284          & 0.246          & 0.248          & \multicolumn{1}{c|}{0.236}          & 0.385           & 0.302           & 0.751          & 0.778          & \multicolumn{1}{c|}{0.748}          & 0.914          \\
                                           & Multi-Fed-Avg           & 0.296          & 0.299          & 0.289          & 0.250          & 0.254          & \multicolumn{1}{c|}{0.245}          & -               & -               & 0.757          & 0.791          & \multicolumn{1}{c|}{0.759}          & -              \\
                                           & FedMLLM                 & 0.295          & 0.301          & 0.291          & 0.252          & 0.257          & \multicolumn{1}{c|}{0.242}          & -               & -               & 0.782          & 0.811          & \multicolumn{1}{c|}{0.756}          & -              \\
                                           & Fedilora                & 0.299          & 0.304          & 0.291          & 0.253          & 0.259          & \multicolumn{1}{c|}{0.246}          & -               & -               & 0.794          & 0.818          & \multicolumn{1}{c|}{0.772}          & -              \\
                                           & Co-PLMs                 & 0.303          & 0.308          & 0.301          & 0.258          & 0.261          & \multicolumn{1}{c|}{0.248}          & 0.392           & 0.304           & 0.794          & 0.819          & \multicolumn{1}{c|}{0.762}          & 0.925          \\
                                           & \textbf{Ours}           & \textbf{0.309} & \textbf{0.313} & \textbf{0.301} & \textbf{0.262} & \textbf{0.264} & \multicolumn{1}{c|}{\textbf{0.251}} & \textbf{0.398}  & \textbf{0.311}  & \textbf{0.801} & \textbf{0.824} & \multicolumn{1}{c|}{\textbf{0.771}} & \textbf{0.931} \\ \hline
\end{tabular}
}
\begin{tablenotes}
\footnotesize
\item “–”: Multi-Fed-Avg, FedMLLM and Fedilora are not applicable for evaluating server performance.
\end{tablenotes}
\end{threeparttable}
\end{table*}

\subsection{Communication Cost Analysis (RQ2)}
\begin{figure}[H]
\vspace{-1mm}
    \centering
    \includegraphics[width=0.8\linewidth]{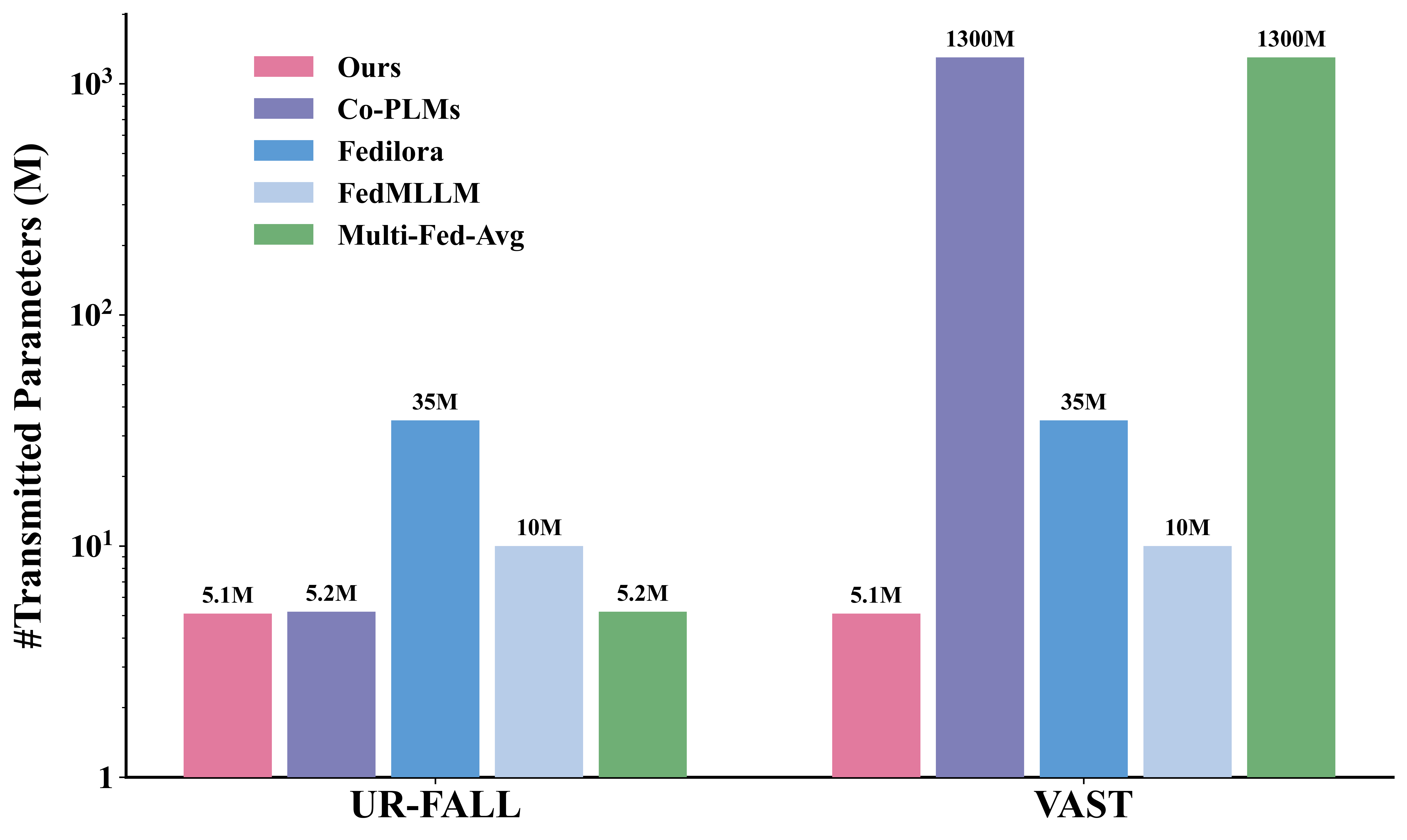}
    \vspace{-5mm}
  \caption{Communication cost.}
  \label{fig:3}
  \vspace{-5mm}
\end{figure}
To evaluate communication efficiency, we compare the baselines based on the average number of parameters transmitted by each edge device per communication round, as illustrated in \figurename~\ref{fig:3}. In the edge-cloud synergy setting, all edge devices employ the same SLM backbone while using different multimodal feature encoders according to their available modalities. Under this setting, \pname achieves superior communication efficiency by transmitting only the low-rank LoRA parameters and fused multimodal representations. Each edge device transmits an average of 5.1 million parameters per communication round, accounting for only 0.71\% of the parameters in the local model. In contrast, Multi-FedAvg requires the transmission of the full set of trained encoder parameters, resulting in prohibitively high communication cost. Similarly, Co-PLMs exchanges both LoRA adapter and encoder parameters, leading to substantially larger transmission volumes. Although Fedilora and FedMLLM also employ parameter-efficient fine-tuning by communicating only LoRA parameters, their communication costs remain considerably higher than those of \pname. Specifically, Fedilora adopts a higher LoRA rank ($r=24$) than \pname ($r=8$), substantially increasing the number of transmitted adapter parameters. Consequently, its communication cost is approximately \textbf{seven times} that of \pname. Although FedMLLM selectively updates intermediate LoRA layers, it requires the transmission of additional auxiliary parameters, resulting in a communication cost approximately \textbf{twice} that of \pname. Overall, \pname achieves substantially lower communication cost while maintaining competitive model performance.
\begin{table}[t!]
\vspace{-3mm}
\caption{Performance comparison of different number of edge devices}
\vspace{-3mm}
\label{tab:scala}
\resizebox{0.85\linewidth}{!}{
\begin{tabular}{|ccc|cccc|}
\hline
\multicolumn{3}{|c|}{\multirow{2}{*}{}}                                                                        & \multicolumn{4}{c|}{Number of Edge Devices}                                                  \\ \cline{4-7} 
\multicolumn{3}{|c|}{}                                                                                         & \multicolumn{1}{c|}{3}     & \multicolumn{1}{c|}{5}     & \multicolumn{1}{c|}{10}    & 20    \\ \hline
\multicolumn{1}{|c|}{\multirow{8}{*}{VAST}}    & \multicolumn{1}{c|}{\multirow{6}{*}{Client Perf.}} & Avg. BS  & \multicolumn{1}{c|}{0.309} & \multicolumn{1}{c|}{0.308} & \multicolumn{1}{c|}{0.302} & 0.298 \\ \cline{3-7} 
\multicolumn{1}{|c|}{}                         & \multicolumn{1}{c|}{}                              & B. BS    & \multicolumn{1}{c|}{0.313} & \multicolumn{1}{c|}{0.314} & \multicolumn{1}{c|}{0.317} & 0.312 \\ \cline{3-7} 
\multicolumn{1}{|c|}{}                         & \multicolumn{1}{c|}{}                              & W. BS    & \multicolumn{1}{c|}{0.301} & \multicolumn{1}{c|}{0.298} & \multicolumn{1}{c|}{0.294} & 0.288 \\ \cline{3-7} 
\multicolumn{1}{|c|}{}                         & \multicolumn{1}{c|}{}                              & Avg. RLS & \multicolumn{1}{c|}{0.262} & \multicolumn{1}{c|}{0.261} & \multicolumn{1}{c|}{0.256} & 0.254 \\ \cline{3-7} 
\multicolumn{1}{|c|}{}                         & \multicolumn{1}{c|}{}                              & B. RLS   & \multicolumn{1}{c|}{0.264} & \multicolumn{1}{c|}{0.264} & \multicolumn{1}{c|}{0.261} & 0.259 \\ \cline{3-7} 
\multicolumn{1}{|c|}{}                         & \multicolumn{1}{c|}{}                              & W. RLS   & \multicolumn{1}{c|}{0.251} & \multicolumn{1}{c|}{0.253} & \multicolumn{1}{c|}{0.247} & 0.240 \\ \cline{2-7} 
\multicolumn{1}{|c|}{}                         & \multicolumn{1}{c|}{\multirow{2}{*}{Server Perf.}} & BS       & \multicolumn{1}{c|}{0.398} & \multicolumn{1}{c|}{0.396} & \multicolumn{1}{c|}{0.398} & 0.399 \\ \cline{3-7} 
\multicolumn{1}{|c|}{}                         & \multicolumn{1}{c|}{}                              & RLS      & \multicolumn{1}{c|}{0.311} & \multicolumn{1}{c|}{0.308} & \multicolumn{1}{c|}{0.310} & 0.311 \\ \hline\hline
\multicolumn{1}{|c|}{\multirow{4}{*}{UR-FALL}} & \multicolumn{1}{c|}{\multirow{3}{*}{Client Perf.}} & Avg. F1  & \multicolumn{1}{c|}{0.801} & \multicolumn{1}{c|}{0.803} & \multicolumn{1}{c|}{0.796} & 0.794 \\ \cline{3-7} 
\multicolumn{1}{|c|}{}                         & \multicolumn{1}{c|}{}                              & B. F1    & \multicolumn{1}{c|}{0.824} & \multicolumn{1}{c|}{0.828} & \multicolumn{1}{c|}{0.827} & 0.823 \\ \cline{3-7} 
\multicolumn{1}{|c|}{}                         & \multicolumn{1}{c|}{}                              & W. F1    & \multicolumn{1}{c|}{0.771} & \multicolumn{1}{c|}{0.768} & \multicolumn{1}{c|}{0.752} & 0.754 \\ \cline{2-7} 
\multicolumn{1}{|c|}{}                         & \multicolumn{1}{c|}{Server Perf.}                  & F1       & \multicolumn{1}{c|}{0.931} & \multicolumn{1}{c|}{0.930} & \multicolumn{1}{c|}{0.927} & 0.928 \\ \hline
\end{tabular}
}
\vspace{-5mm}
\end{table}
\subsection{Scalability Analysis (RQ3)}
We investigate the scalability of \pname by evaluating its predictive performance as the number of edge devices increases. As shown in Table~\ref{tab:scala}, we vary the number of edge devices in the edge-cloud synergy over $\{3, 5, 10, 20\}$ and report the corresponding device-side and server-side performance on the VAST and UR-FALL datasets. On the VAST dataset, the device-side Avg. BS decreases slightly from 0.309 to 0.298, while the Avg. RLS declines from 0.262 to 0.254 as the number of edge devices increases from 3 to 20. On the UR-FALL dataset, the device-side Avg. F1 score decreases modestly from 0.801 to 0.794, with a reduction of only 0.87\%. In contrast, the server-side performance remains stable as the number of edge devices increases. Specifically, on the VAST dataset, the server-side BS and RLS remain within the narrow ranges of 0.396--0.399 and 0.308--0.311, respectively. On the UR-FALL dataset, the server-side F1 score decreases only marginally from 0.931 to 0.928. These results demonstrate that \pname maintains robust model performance as the number of edge devices increases during the collaborative learning process. We also investigate the sensitivity of \pname to the LoRA rank, with a detailed analysis provided in \textbf{Appendix C}.

\subsection{Impact of Edge Model Size (RQ4)}

\begin{figure}[t]
    \centering
        \vspace{-5mm}
\includegraphics[width=0.75\columnwidth]{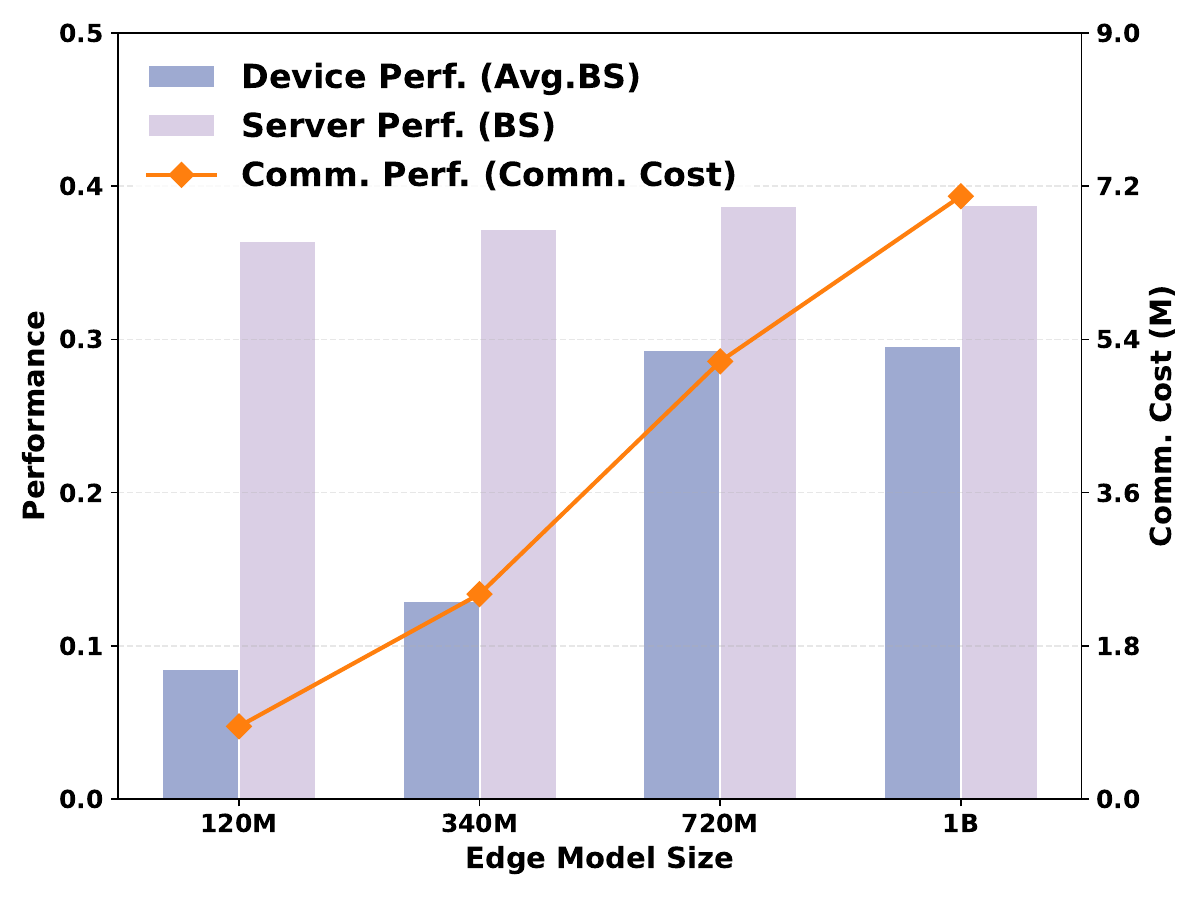}
    \vspace{-5mm}
    \caption{Impact of edge model size (120M, 340M, 720M, 1B) on device-side performance, server-side performance, and communication performance on the VAST dataset.}
    \label{fig:impact_of_model_size}
    \vspace{-5mm}
\end{figure}
\begin{table*}[t!]
\caption{Ablation study of \pname on the VAST and UR-FALL datasets.}
\vspace{-3mm}
\centering
\label{tab:abl}
\resizebox{0.85\textwidth}{!}{\begin{tabular}{|l|cccccccc|cccc|}
\hline
\multirow{3}{*}{\textbf{Model}} & \multicolumn{8}{c|}{\textbf{VAST}}                                                                                                                                                                                                                                                         & \multicolumn{4}{c|}{\textbf{UR-FALL}}                                                                                                     \\ \cline{2-13} 
                                & \multicolumn{6}{c|}{\textbf{Device Perf.}}                                                                                                                                                                                               & \multicolumn{2}{c|}{\textbf{Server Perf.}}      & \multicolumn{3}{c|}{\textbf{Device Perf.}}                                                                        & \textbf{Server Perf.} \\ \cline{2-13} 
                                & \multicolumn{1}{c|}{\textbf{Avg. BS}} & \multicolumn{1}{c|}{\textbf{B. BS}} & \multicolumn{1}{c|}{\textbf{W. BS}} & \multicolumn{1}{c|}{\textbf{Avg. RLS}} & \multicolumn{1}{c|}{\textbf{B. RLS}} & \multicolumn{1}{c|}{\textbf{W. RLS}} & \multicolumn{1}{c|}{\textbf{BS}} & \textbf{RLS} & \multicolumn{1}{c|}{\textbf{Avg. F1}} & \multicolumn{1}{c|}{\textbf{B. F1}} & \multicolumn{1}{c|}{\textbf{W. F1}} & \textbf{F1}           \\ \hline
\textbf{Ours}                   & \multicolumn{1}{c|}{0.293}            & \multicolumn{1}{c|}{0.301}          & \multicolumn{1}{c|}{0.284}          & \multicolumn{1}{c|}{0.244}             & \multicolumn{1}{c|}{0.249}           & \multicolumn{1}{c|}{0.232}           & \multicolumn{1}{c|}{0.387}       & 0.304        & \multicolumn{1}{c|}{0.754}            & \multicolumn{1}{c|}{0.791}          & \multicolumn{1}{c|}{0.698}          & 0.918                 \\ \hline
\textbf{-w/o $\text{CCL}_c$}    & \multicolumn{1}{c|}{0.266}            & \multicolumn{1}{c|}{0.274}          & \multicolumn{1}{c|}{0.240}          & \multicolumn{1}{c|}{0.221}             & \multicolumn{1}{c|}{0.239}           & \multicolumn{1}{c|}{0.206}           & \multicolumn{1}{c|}{0.371}       & 0.278        & \multicolumn{1}{c|}{0.725}            & \multicolumn{1}{c|}{0.749}          & \multicolumn{1}{c|}{0.678}          & 0.890                 \\ \hline
\textbf{-w/o $\text{CCL}_i$}    & \multicolumn{1}{c|}{0.270}            & \multicolumn{1}{c|}{0.276}          & \multicolumn{1}{c|}{0.243}          & \multicolumn{1}{c|}{0.222}             & \multicolumn{1}{c|}{0.238}           & \multicolumn{1}{c|}{0.213}           & \multicolumn{1}{c|}{0.373}       & 0.279        & \multicolumn{1}{c|}{0.731}            & \multicolumn{1}{c|}{0.754}          & \multicolumn{1}{c|}{0.680}          & 0.894                 \\ \hline
\textbf{-w/o $\text{MAA}_a$}    & \multicolumn{1}{c|}{0.275}            & \multicolumn{1}{c|}{0.279}          & \multicolumn{1}{c|}{0.266}          & \multicolumn{1}{c|}{0.229}             & \multicolumn{1}{c|}{0.238}           & \multicolumn{1}{c|}{0.218}           & \multicolumn{1}{c|}{0.367}       & 0.285        & \multicolumn{1}{c|}{0.732}            & \multicolumn{1}{c|}{0.778}          & \multicolumn{1}{c|}{0.674}          & 0.889                 \\ \hline
\textbf{-w/o $\text{MAA}_g$}    & \multicolumn{1}{c|}{0.278}            & \multicolumn{1}{c|}{0.286}          & \multicolumn{1}{c|}{0.269}          & \multicolumn{1}{c|}{0.231}             & \multicolumn{1}{c|}{0.243}           & \multicolumn{1}{c|}{0.217}           & \multicolumn{1}{c|}{0.367}       & 0.286        & \multicolumn{1}{c|}{0.735}            & \multicolumn{1}{c|}{0.769}          & \multicolumn{1}{c|}{0.689}          & 0.887                 \\ \hline
\textbf{-w/o SE-CCL}            & \multicolumn{1}{c|}{0.269}            & \multicolumn{1}{c|}{0.280}          & \multicolumn{1}{c|}{0.257}          & \multicolumn{1}{c|}{0.219}             & \multicolumn{1}{c|}{0.223}           & \multicolumn{1}{c|}{0.211}           & \multicolumn{1}{c|}{0.383}       & 0.299        & \multicolumn{1}{c|}{0.698}            & \multicolumn{1}{c|}{0.735}          & \multicolumn{1}{c|}{0.631}          & 0.876                 \\ \hline
\end{tabular}
}
\end{table*}

    
    

We further investigate the impact of the edge model size on both model performance and communication cost. Specifically, we evaluate \pname using edge models of different sizes, comprising 120M, 340M, 720M, and 1B parameters, initialized following prior work~\cite{gu2023minillm}. As shown in Figure 4, increasing the edge model size from 120M to 720M substantially improves the performance of \pname on both datasets. On the VAST dataset, the device-side Avg. BS increases from 0.085 to 0.293, corresponding to an improvement of 244.71\%.  Although the server-side BS increases from 0.364 to 0.387, representing a 6.32\% improvement, this gain is substantially smaller than those observed in the device-side metrics. It is worth noting that further increasing the model size from 720M to 1B yields only marginal performance gains. Specifically, the average device-side Avg. BS on the VAST dataset improves by only 1.02\%. In contrast, the communication cost increases substantially: from 5.142M to 7.083M parameters on the VAST dataset, representing increases of 37.75\%. Accordingly, we adopt the 720M model as the default edge model, as it offers a favorable trade-off between predictive performance and communication cost. Detailed experimental results on both the VAST and UR-FALL datasets are provided in \textbf{Appendix D}.

\subsection{Ablation Study (RQ5)}

To validate the contributions of the key components in \pname, we construct five variants by removing the corresponding components, namely \pname w/o $\text{CCL}_{c}$, \pname w/o $\text{CCL}_{i}$, $\text{\pname}\;\text{w/o}\;\text{MAA}_{a}$, $\text{\pname}\;\text{w/o}\;\text{MAA}_{g}$, and $\text{\pname}\;\text{w/o}\;\text{SE\mbox{-}CCL}$. Table~\ref{tab:abl} reports the device-side and server-side results on the VAST and UR-FALL datasets. Overall, \pname consistently outperforms all variants by 2.59\%--11.42\% in terms of device-side  performance across both datasets, demonstrating the contribution of each component to the effectiveness of the proposed framework. The detailed figures illustrating the results are provided in \textbf{Appendix E}.

\begin{sloppypar}
\pname w/o $\text{CCL}_{c}$ and \pname w/o $\text{CCL}_{i}$ variants replace the vector-volume-based method with cosine similarity and InfoNCE~\cite{liu2024graph}, respectively, to construct the cross-modal contrastive loss. On the VAST dataset, removing CCL leads to a clear performance decline. Compared with \pname w/o $\text{CCL}_{c}$ and \pname w/o $\text{CCL}_{i}$, \pname improves the device-side Avg. BS by 10.15\% and 8.52\%, respectively, and the device-side Avg. RLS by 10.41\% and 9.91\%, respectively. The server-side BS and RLS exhibit similar trends. Similarly, on the UR-FALL dataset, \pname achieves average gains of 3.57\% and 2.92\% in Avg. F1 and F1, respectively. These results indicate that the vector-volume-based method in CCL improves multimodal representation learning within the edge-cloud synergy.
\end{sloppypar}

To investigate the effectiveness of MAA, \pname w/o $\text{MAA}_{a}$ and \pname w/o $\text{MAA}_{g}$ variants remove the modality-aware aggregation mechanism at the server and instead aggregate updates from all edge devices using simple averaging and a gradient-similarity-based mechanism~\cite{liu2025towards}, respectively. As shown in Table~\ref{tab:abl}, \pname outperforms \pname w/o $\text{MAA}_{a}$ and \pname w/o $\text{MAA}_{g}$ variants by an average of 5.10\% in terms of server-side performance, with BS and RLS used as the evaluation metrics on the VAST dataset and F1 on the UR-FALL dataset. The improvements in both device-side and server-side performance can be attributed to MAA, which adaptively assigns aggregation weights according to modality availability across edge devices, thereby mitigating the performance degradation caused by modality heterogeneity.

Apart from that, we construct \pname w/o $\text{SE\mbox{-}CCL}$ variant by removing the SE-CCL process, retaining only local training on the edge devices and server-side aggregation, to evaluate the effectiveness of SE-CCL. As shown in Table~\ref{tab:abl}, \pname outperforms this variant by 8.92\%, 11.42\%, 1.04\%, and 1.67\% in Avg. BS, Avg. RLS, BS, and RLS, respectively, on the VAST dataset. A similar trend is observed for the device-side Avg. F1 and server-side F1 on the UR-FALL dataset. These results underscore the importance of SE-CCL, enabling knowledge transfer between on-device and server-based models, effectively enhancing the model performance of models within edge-cloud synergy.

\section{Conclusion}
In this paper, we propose \pname, a collaborative multimodal learning framework for edge-cloud synergy that jointly trains a server-side model and domain-specific edge models. \pname improves both edge and server performance while preserving data privacy under modality and model-structure heterogeneity. In each communication round, the server generates fused omnimodal representations using a unified model and distributes them to edge devices. Each device then performs CCL and AMT, and uploads the trainable LoRA parameters of its SLM backbone. The server aggregates these parameters via MAA to update the server-side SLM, applies SE-CCL, and distributes the updated LoRA parameters back to the devices for local SLM updates. Experiments on both question answering and classification tasks show that \pname outperforms state-of-the-art baselines while maintaining communication efficiency. Future work will extend \pname to dynamic edge-cloud environments with time-varying modality availability while maintaining performance robustness.


\bibliographystyle{ACM-Reference-Format}
\bibliography{reference}
\newpage
\nobalance
\appendix
\section{Pseudocode of \pname}
The collaborative training process of \pname is described in Algorithm~\ref{algo}.
\renewcommand{\thealgorithm}{1}
\begin{algorithm}
\caption{\pname}\label{algo:1}
\begin{algorithmic}[1]
\Require The set of devices $\mathcal{D}$, the server, the number of device $N$, the number of communication round $T$, the public dataset $D^{'}$, the set of private dataset across devices $\{D_j\}_{j = 1}^{N}$, the set of public dataset across devices $\{D^{'}_j\}_{j = 1}^{N}$, the server-based unified model $\emph{M}^s$, the server-based SLM $\emph{B}^s_{slm}$ and the set of unified models across devices $\{\emph{M}^d_j\}_{j = 1}^{N}$
\For{each round $t\in [1,2,...,T]$}
\State // Server side.
\State Generate the fused omnimodal representation $\emph{s}^{'}(t)$ by Eq.~\ref{eq:fused} on the public dataset $D^{'}$ and distribute $\emph{s}^{'}(t)$ to each device
\State // Device side.
\For{each device $d_j\in\mathcal{D}$ (in parallel)}
\State$\Bar{\emph{M}}_{j}^{d}(t)\leftarrow f_{ccl}(\emph{M}_j^{d}(t);D^{'}_j)$\Comment{Cross-modal contrastive learning}
\State $\hat{\emph{M}}_{j}^{d}(t),\phi_{lora}(\emph{B}^d_j(t))\leftarrow f_{amt}(\Bar{\emph{M}}_{j}^{s}(t);D_j)$ \Comment{Adaptive multimodal tuning}
\State upload $\phi_{lora}(\emph{B}^d_j(t))$ and $[\mathcal{M}_j]$ to the server
\EndFor
\State // Server side.
\State Assign aggregation weights by Eq.\ref{eq:weight}
\State $\phi_{lora}(B^s_{slm}(t))\leftarrow f_{maa}(\{\emph{B}^d_{j}(t)\}_{j = 1}^{N})$ \Comment{Modality-aware model aggregation}
\State Update $\emph{B}^{s}_{slm}(t)$ by $\phi_{lora}(B^s_{slm}(t))$  
\State  $\emph{M}_{s}(t+1), \phi_{lora}(\emph{B}^{s}_{slm}(t+1))\leftarrow f_{se}(\emph{M}^{s}(t), \emph{B}^{s}_{slm}(t))$ \Comment{SLM-enhanced cross-modal contrastive learning}
\State Distribute $\phi_{lora}(\emph{B}^{s}_{slm}(t+1))$ to each device
\State // Device side.
\For{each device $d_j\in\mathcal{D}$ (in parallel)}
\State $\phi_{lora}(\emph{B}^{d}_j(t+1))\leftarrow\phi_{lora}(\emph{B}^{s}_{slm}(t+1))$
\State Update $\emph{B}^{d}_j(t+1)$ by $\phi_{lora}(\emph{B}^{d}_j(t+1))$
\EndFor
\EndFor
\end{algorithmic}
\label{algo}
\end{algorithm}

\section{Baseline Details}

We conduct a comparative analysis of \pname against the following baselines:
\begin{itemize}[leftmargin=0.3cm, noitemsep]
\item\textbf{Standalone} allows both edge devices and the server to independently train their local models using their own datasets without any collaborative learning.

\item\textbf{Multi-Fed-Avg} extends vanilla FedAvg~\cite{li2019convergence} to our multimodal federated datasets where edge devices collaboratively train multi-modality models.

\item\textbf{FedMLLM}~\cite{xu2024fedmllm} applies lightweight quantized MLLMs with LoRA for efficient device-side training, and introduces two modality-agnostic strategies: prompt-based debiasing that explicitly instructs models to ignore modality information, and adaptive layer-wise regularization that dynamically adjusts constraint strength based on missing modality rates and selectively regularizes intermediate layers to preserve modality-specific and task-specific features.

\item\textbf{Fedilora}~\cite{yang2025fedilora} employs a dimension-wise reweighting strategy that aggregates LoRA parameters without zero-padding dilution, and a layer-wise model editing technique that repairs local LoRA layers using global parameters based on cosine similarity, addressing both heterogeneous LoRA ranks and missing modalities in federated multimodal learning.

\item\textbf{Co-PLMs}~\cite{liu2025structure} employs distilled proxy models (DPMs) as bridges to enable bidirectional knowledge transfer between heterogeneous on-device SLMs and the server-based LLM through structure-agnostic mutual learning (SAML), with domain-specific tuning (DST) preserving local domain knowledge and SAML addressing tokenizer mismatch and logit divergence issues.
\end{itemize}

\section{Impact of LoRA Rank $r$}
\begin{table}[!t]
\centering
\caption{Performance and communication cost of \pname under different LoRA ranks.}
\label{tab:rank}

\begin{minipage}{\linewidth}
\centering
\begin{threeparttable}

\resizebox{1.02\linewidth}{!}{
\begin{tabular}{|ccc|ccccc|}
\hline
\multicolumn{3}{|c|}{\multirow{2}{*}{}} 
& \multicolumn{5}{c|}{LoRA Rank $r$} \\ 
\cline{4-8}

\multicolumn{3}{|c|}{} 
& \multicolumn{1}{c|}{2} 
& \multicolumn{1}{c|}{4} 
& \multicolumn{1}{c|}{\textbf{8}} 
& \multicolumn{1}{c|}{16} 
& 24 \\ 
\hline

\multicolumn{1}{|c|}{\multirow{9}{*}{VAST}} 
& \multicolumn{1}{c|}{\multirow{6}{*}{Device Perf. $\uparrow$}} 
& Avg. BS 
& \multicolumn{1}{c|}{0.167} 
& \multicolumn{1}{c|}{0.214} 
& \multicolumn{1}{c|}{0.293} 
& \multicolumn{1}{c|}{0.296} 
& 0.296 \\ 
\cline{3-8}

\multicolumn{1}{|c|}{} 
& \multicolumn{1}{c|}{} 
& B. BS 
& \multicolumn{1}{c|}{0.191} 
& \multicolumn{1}{c|}{0.265} 
& \multicolumn{1}{c|}{0.301} 
& \multicolumn{1}{c|}{0.302} 
& 0.303 \\ 
\cline{3-8}

\multicolumn{1}{|c|}{} 
& \multicolumn{1}{c|}{} 
& W. BS 
& \multicolumn{1}{c|}{0.120} 
& \multicolumn{1}{c|}{0.186} 
& \multicolumn{1}{c|}{0.284} 
& \multicolumn{1}{c|}{0.287} 
& 0.281 \\ 
\cline{3-8}

\multicolumn{1}{|c|}{} 
& \multicolumn{1}{c|}{} 
& Avg. RLS 
& \multicolumn{1}{c|}{0.148} 
& \multicolumn{1}{c|}{0.214} 
& \multicolumn{1}{c|}{0.244} 
& \multicolumn{1}{c|}{0.248} 
& 0.247 \\ 
\cline{3-8}

\multicolumn{1}{|c|}{} 
& \multicolumn{1}{c|}{} 
& B. RLS 
& \multicolumn{1}{c|}{0.169} 
& \multicolumn{1}{c|}{0.230} 
& \multicolumn{1}{c|}{0.249} 
& \multicolumn{1}{c|}{0.252} 
& 0.253 \\ 
\cline{3-8}

\multicolumn{1}{|c|}{} 
& \multicolumn{1}{c|}{} 
& W. RLS 
& \multicolumn{1}{c|}{0.118} 
& \multicolumn{1}{c|}{0.171} 
& \multicolumn{1}{c|}{0.232} 
& \multicolumn{1}{c|}{0.229} 
& 0.230 \\ 
\cline{2-8}

\multicolumn{1}{|c|}{} 
& \multicolumn{1}{c|}{\multirow{2}{*}{Server Perf. $\uparrow$}} 
& BS 
& \multicolumn{1}{c|}{0.295} 
& \multicolumn{1}{c|}{0.342} 
& \multicolumn{1}{c|}{0.387} 
& \multicolumn{1}{c|}{0.391} 
& 0.390 \\ 
\cline{3-8}

\multicolumn{1}{|c|}{} 
& \multicolumn{1}{c|}{} 
& RLS 
& \multicolumn{1}{c|}{0.243} 
& \multicolumn{1}{c|}{0.277} 
& \multicolumn{1}{c|}{0.304} 
& \multicolumn{1}{c|}{0.303} 
& 0.306 \\ 
\cline{2-8}

\multicolumn{1}{|c|}{} 
& \multicolumn{1}{c|}{Comm. Perf. $\downarrow$} 
& Comm. Cost (M) 
& \multicolumn{1}{c|}{1.697} 
& \multicolumn{1}{c|}{2.846} 
& \multicolumn{1}{c|}{5.142} 
& \multicolumn{1}{c|}{9.768} 
& 14.342 \\ 
\hline
\hline

\multicolumn{1}{|c|}{\multirow{5}{*}{UR-FALL}} 
& \multicolumn{1}{c|}{\multirow{3}{*}{Device Perf. $\uparrow$}} 
& Avg. F1 
& \multicolumn{1}{c|}{0.592} 
& \multicolumn{1}{c|}{0.683} 
& \multicolumn{1}{c|}{0.754} 
& \multicolumn{1}{c|}{0.753} 
& 0.758 \\ 
\cline{3-8}

\multicolumn{1}{|c|}{} 
& \multicolumn{1}{c|}{} 
& B. F1 
& \multicolumn{1}{c|}{0.624} 
& \multicolumn{1}{c|}{0.719} 
& \multicolumn{1}{c|}{0.791} 
& \multicolumn{1}{c|}{0.791} 
& 0.797 \\ 
\cline{3-8}

\multicolumn{1}{|c|}{} 
& \multicolumn{1}{c|}{} 
& W. F1 
& \multicolumn{1}{c|}{0.555} 
& \multicolumn{1}{c|}{0.657} 
& \multicolumn{1}{c|}{0.698} 
& \multicolumn{1}{c|}{0.703} 
& 0.710 \\ 
\cline{2-8}

\multicolumn{1}{|c|}{} 
& \multicolumn{1}{c|}{Server Perf. $\uparrow$} 
& F1 
& \multicolumn{1}{c|}{0.796} 
& \multicolumn{1}{c|}{0.838} 
& \multicolumn{1}{c|}{0.918} 
& \multicolumn{1}{c|}{0.923} 
& 0.919 \\ 
\cline{2-8}

\multicolumn{1}{|c|}{} 
& \multicolumn{1}{c|}{Comm. Perf. $\downarrow$} 
& Comm. Cost (M) 
& \multicolumn{1}{c|}{1.677} 
& \multicolumn{1}{c|}{2.813} 
& \multicolumn{1}{c|}{5.127} 
& \multicolumn{1}{c|}{9.742} 
& 14.189 \\ 
\hline

\end{tabular}
}

\begin{tablenotes}[flushleft]
\footnotesize
\item[] $\uparrow$: higher is better; $\downarrow$: lower is better.
\end{tablenotes}

\end{threeparttable}
\end{minipage}
\end{table}

To investigate the impact of the LoRA rank, we evaluate the performance and communication cost of \pname for $r \in \{2, 4, 8, 16, 24\}$. Table~\ref{tab:rank} reports the device-side and server-side performance on the VAST and UR-FALL datasets, together with the corresponding communication costs. A larger LoRA rank increases the number of trainable parameters, allowing the model to capture richer task-specific information. However, it also increases the communication cost associated with exchanging model updates between the edge devices and the server.

As shown in Table~\ref{tab:rank}, increasing the LoRA rank from 2 to 8 consistently improves the performance of \pname on both datasets. On the VAST dataset, the average device-side BS and RLS increase from 0.167 and 0.148 to 0.293 and 0.244, corresponding to improvements of 75.45\% and 64.86\%, respectively. The server-side BS and RLS also improve from 0.295 and 0.243 to 0.387 and $0.304$, respectively. Similarly, on the UR-FALL dataset, the average device-side F1 score increases from 0.592 to 0.758, while the server-side F1 score rises from 0.796 to 0.918. However, it is observed that the device-side and server-side performance improves only marginally when $r$ exceeds 8. Further analysis of Table~\ref{tab:rank} demonstrates that the communication cost continuously increases as $r$ grows. For example, when the rank increases from 8 to 24, the average device-side BS and RLS on the VAST dataset improve by only 1.02\% and 1.23\%, respectively. In contrast, the communication cost grows substantially. On the VAST dataset, the communication cost increases from 5.142M at $r=8$ to 14.342M at $r=24$, representing an increase of $178.92\%$. A similar trend is observed on the UR-FALL dataset, where the communication cost rises from 5.127M to 14.189M. Therefore, we set $r=8$ as the default LoRA rank, as it provides a favorable trade-off between model performance and communication cost. This trend is consistently observed on both the VAST and UR-FALL datasets.

\section{Detailed Experimental Results on the Impact of Edge Model Size}
\begin{figure}[h]
    \centering
    
    \begin{subfigure}[b]{\linewidth}
        \centering
        \includegraphics[width=0.9\linewidth]{image/VAST_Results.pdf}
        \caption{Performance on the VAST dataset.}
    \end{subfigure}
    
    \vspace{0.8em} 
    \begin{subfigure}[b]{\linewidth}
        \centering
        \includegraphics[width=0.9\linewidth]{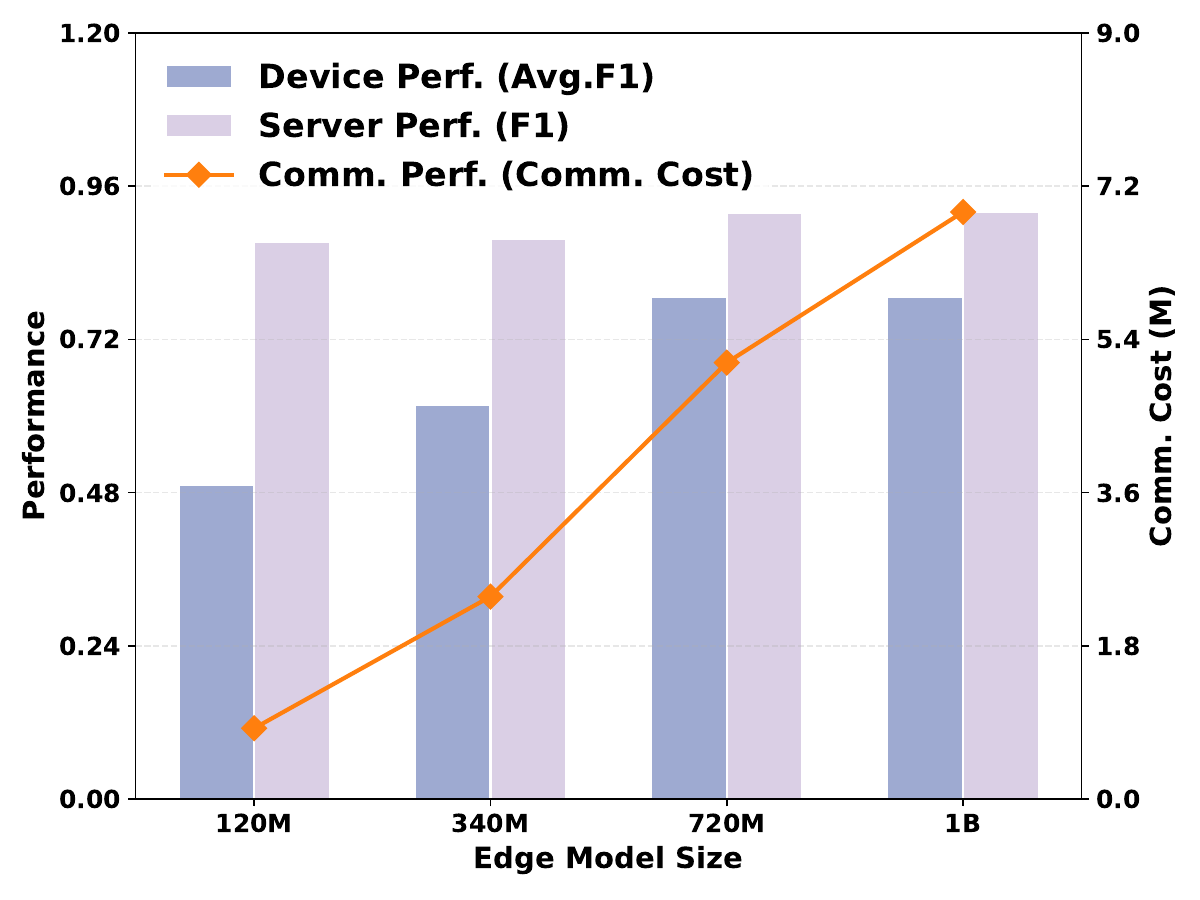}
        \caption{Performance on the UR-FALL dataset.}
    \end{subfigure}
    \caption{Impact of edge model size (120M, 340M, 720M, 1B) on device-side performance, server-side performance, and communication performance.}
    \label{fig:impact_of_model_size}
\end{figure}
\vspace{-3mm}

\section{Detailed Ablation Study Results}
\begin{figure*}[t!]
    \centering
    
    \begin{subfigure}[b]{0.32\textwidth}
        \centering
        \includegraphics[width=\textwidth]{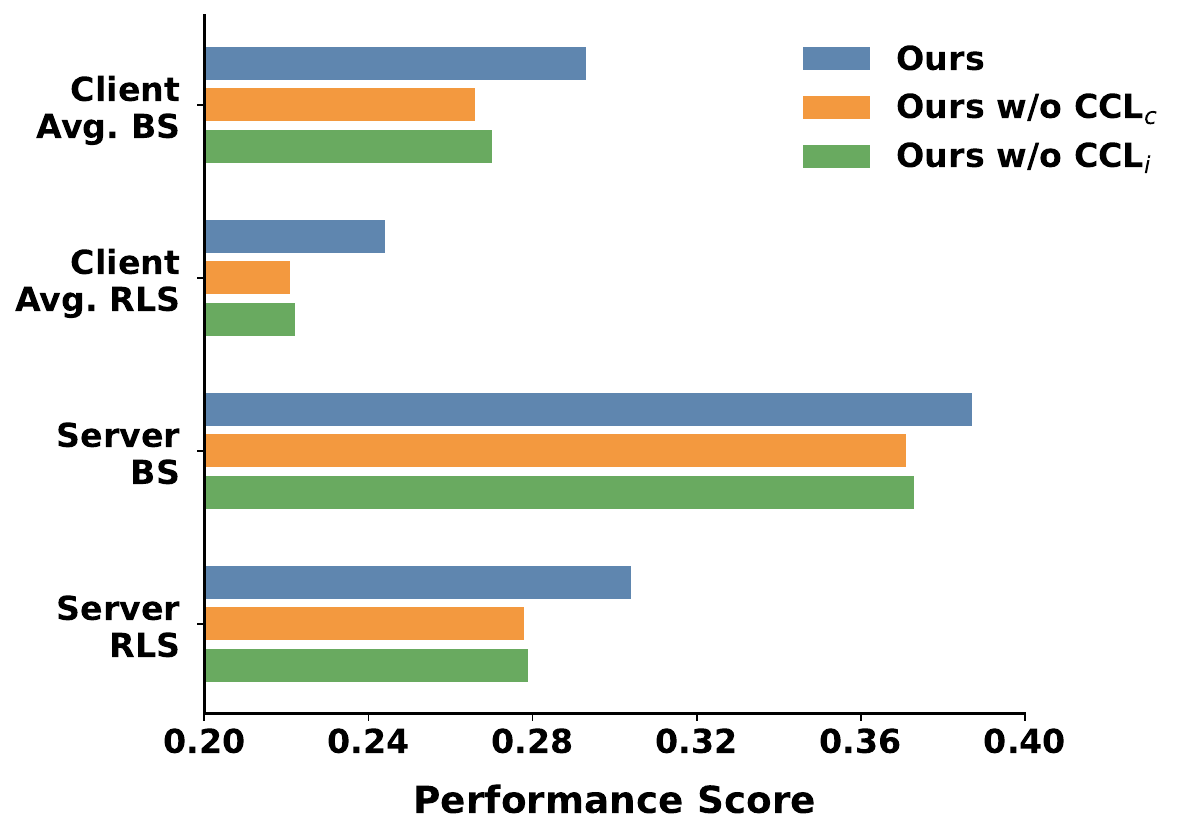}
        \caption{CCL ablation}
        \label{fig:ablation_vast_ccl}
    \end{subfigure}
    \hfill
    \begin{subfigure}[b]{0.32\textwidth}
        \centering
        \includegraphics[width=\textwidth]{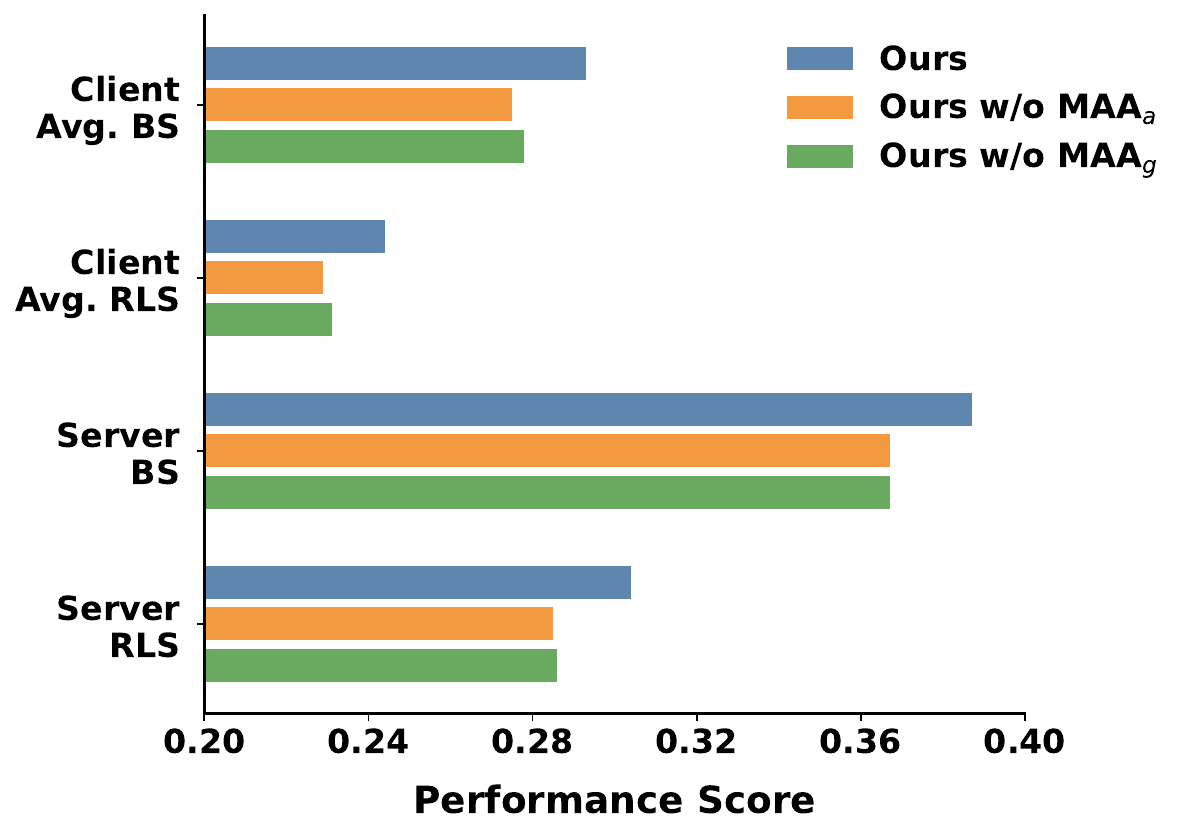}
        \caption{MAA ablation}
        \label{fig:ablation_vast_maa}
    \end{subfigure}%
    \hfill
    \begin{subfigure}[b]{0.32\textwidth}
        \centering
        \includegraphics[width=\textwidth]{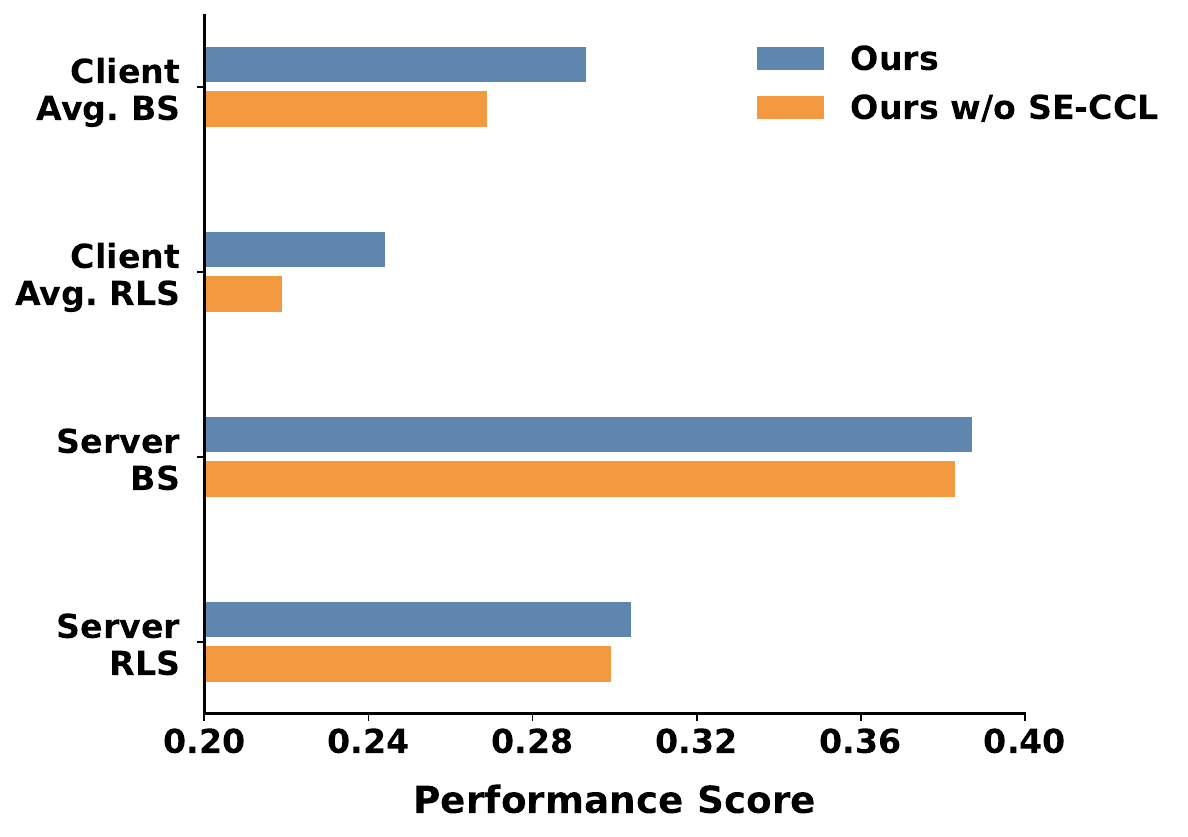}
        \caption{SE-CCL ablation}
        \label{fig:ablation_vast_seccl}
    \end{subfigure}

    \caption{Ablation study results of ML-ECS on the VAST dataset.}
    \label{fig:ablation_study_vast}
\end{figure*}

\begin{figure*}[t]
    \centering
    
    \begin{subfigure}[b]{0.32\textwidth}
        \centering
        \includegraphics[width=\textwidth]{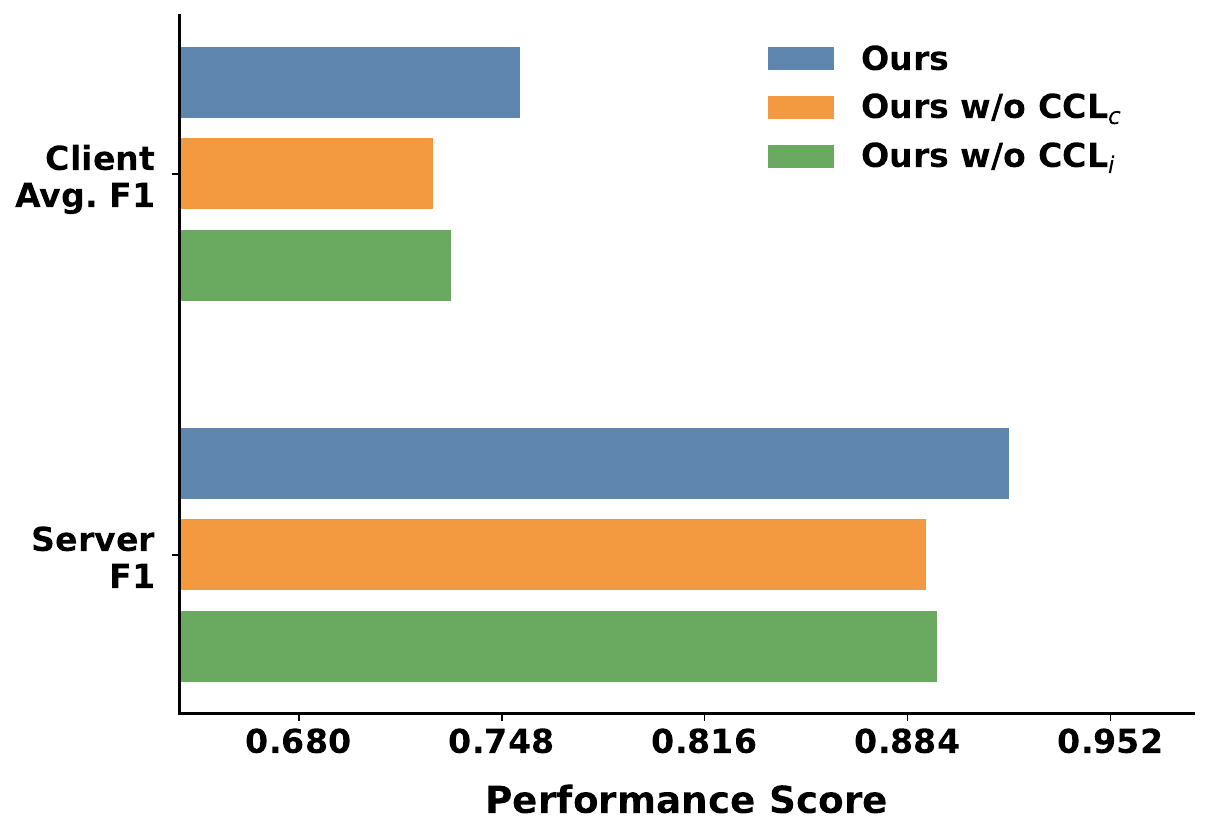}
        \caption{CCL ablation}
        \label{fig:ablation_urfall_ccl}
    \end{subfigure}%
    \hfill
    \begin{subfigure}[b]{0.32\textwidth}
        \centering
        \includegraphics[width=\textwidth]{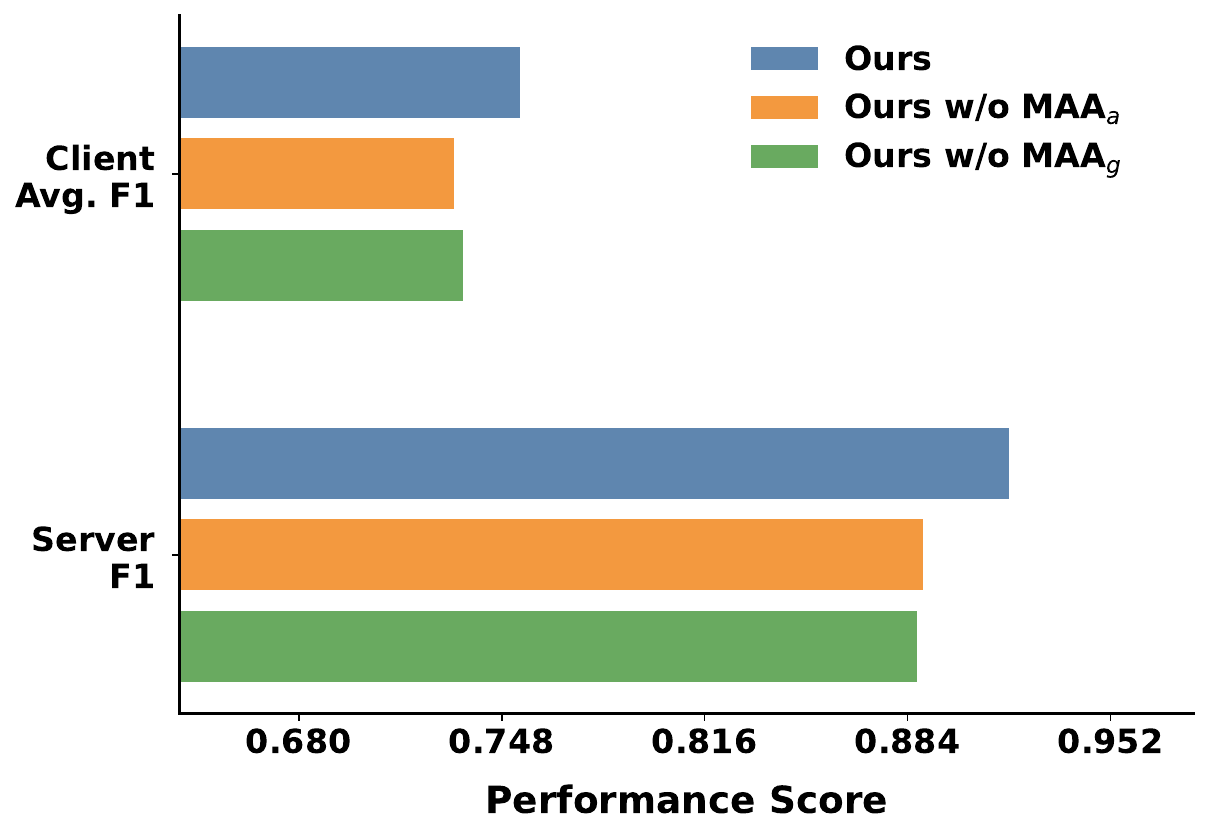}
        \caption{MAA ablation}
        \label{fig:ablation_urfall_maa}
    \end{subfigure}%
    \hfill
    \begin{subfigure}[b]{0.32\textwidth}
        \centering
        \includegraphics[width=\textwidth]{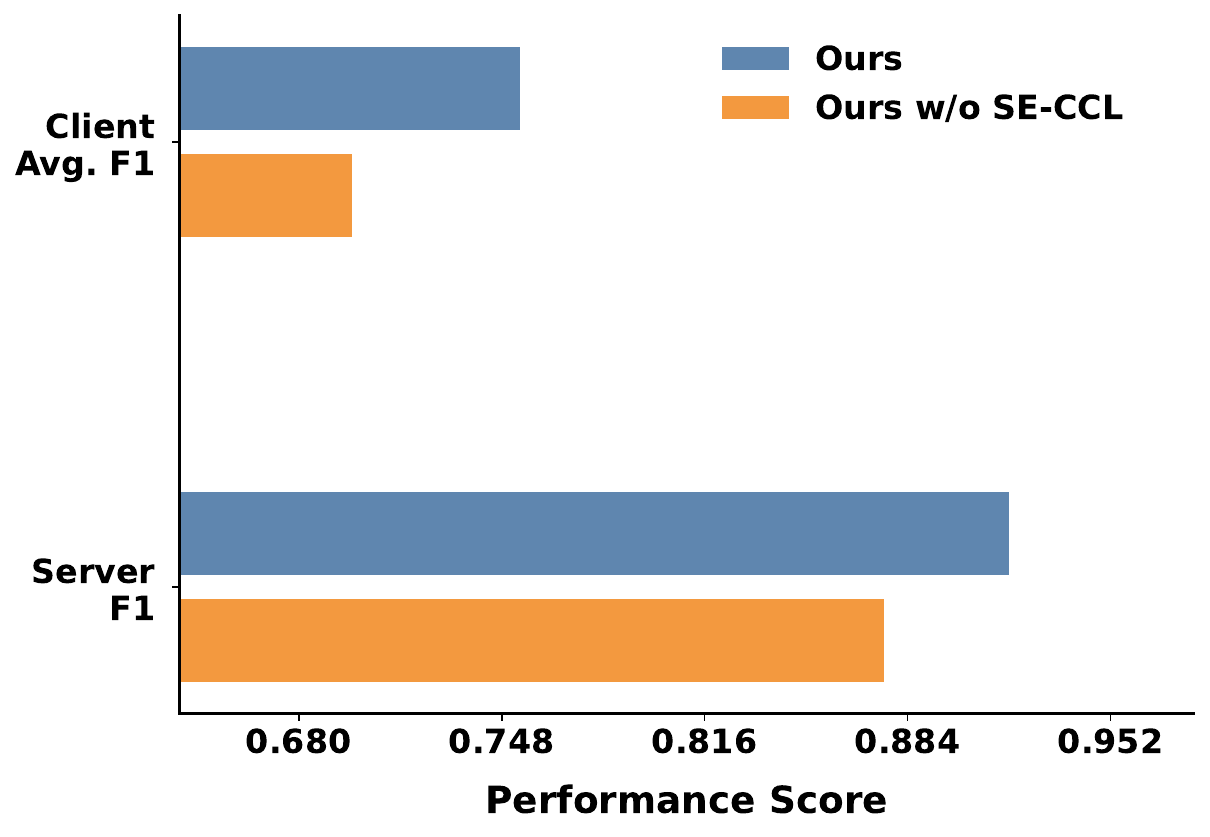}
        \caption{SE-CCL ablation}
        \label{fig:ablation_urfall_seccl}
    \end{subfigure}

    \caption{Ablation study results of ML-ECS on the UR-FALL dataset.}
    \label{fig:ablation_study_urfall}
\end{figure*}
\end{document}